\documentclass[nojo,sglanonrev]{nojournal}
\RequirePackage{tgtermes}
\RequirePackage{newtxtext}
\RequirePackage{newtxmath}
\RequirePackage{endnotes}

\OneAndAHalfSpacedXII 

\usepackage{algorithm}
\usepackage{algpseudocode}
\usepackage{tikz}

\usepackage{natbib}
 \bibpunct[, ]{(}{)}{,}{a}{}{,}%
 \def\bibfont{\small}%

\usepackage{amssymb}
\usepackage{amsmath}
\usepackage{mathtools}

\usepackage{graphicx}
\usepackage{subcaption}
\usepackage{enumitem}
\usepackage{amssymb}
\usepackage{adjustbox}
\usepackage{derivative}

\newcommand{\one}{\ensuremath{\mathbf{1}}}
\newcommand{\zero}{\ensuremath{\mathbf{0}}}

\newcommand{\ind}{\ensuremath{\mathbf{I}}}
\newcommand{\reals}{\ensuremath{\mathbb{R}}}

\DeclareMathOperator{\prob}{\ensuremath{\mathsf{P}}}
\DeclareMathOperator{\expec}{\ensuremath{\mathsf{E}}}

\newcommand{\iso}{\ensuremath{\mathsf{ISO}(\mu_k, h_k)}}
\newcommand{\sys}[1][]{\ensuremath{\mathsf{SYS}_K{#1}}}
\newcommand{\comp}{\ensuremath{[K]}}
\newcommand{\compn}{\ensuremath{\mathsf{K}}}
\newcommand{\uset}{\ensuremath{{U}}}
\newcommand{\sign}{\ensuremath{S}}
\newcommand{\thput}{\ensuremath{\mathsf{Th}}}
\newcommand{\thputD}{\ensuremath{\thput_\mathsf{D}}}
\newcommand{\val}{\ensuremath{\mathsf{Val}}}
\newcommand{\valD}{\ensuremath{\val_\mathsf{D}}}
\newcommand{\fsign}{\ensuremath{\bar{\sign}}}
\newcommand{\exsign}{\ensuremath{\mathcal{S}}}
\usepackage{diffcoeff,amssymb}

\usepackage{xspace}
\usepackage{bbold}

\EquationsNumberedThrough
\TheoremsNumberedThrough 
\ECRepeatTheorems  %

\begin{document}

\RUNAUTHOR{Mirzavand Boroujeni, Iyer, and Cooper}

\RUNTITLE{Decentralized Signaling Mechanisms}

\TITLE{Decentralized Signaling Mechanisms}

\ARTICLEAUTHORS{%
\AUTHOR{Niloufar Mirzavand Boroujeni}
\AFF{Institute for Data Science in Oncology, The University of Texas MD Anderson Cancer Center, \EMAIL{nmirzavand@mdanderson.org}}
\AUTHOR{Krishnamurthy Iyer}
\AFF{Department of Industrial and Systems Engineering, University of Minnesota, \EMAIL{kriyer@umn.edu}}
\AUTHOR{William L. Cooper}
\AFF{Department of Industrial and Systems Engineering, University of Minnesota, \EMAIL{billcoop@umn.edu}}
} 

\ABSTRACT{
We study a system composed of multiple distinct service locations that aims to convince customers to join the system by providing information to customers. 
We cast the system's information design problem in the framework of Bayesian persuasion and describe centralized and decentralized signaling.  We provide efficient methods for computing the system's optimal centralized and decentralized signaling mechanisms and derive a performance guarantee for decentralized signaling when the locations' states are independent. The guarantee states that the probability that a customer joins under optimal decentralized signaling is bounded below by the product of a strictly positive constant and the probability that a customer joins under optimal centralized signaling.  The constant depends only on the number of service locations.  We provide an example that shows that the constant cannot be improved. We consider an extension to more-general objectives for the system and establish that the same guarantee continues to hold. We also extend our analysis to systems where the locations' states are correlated, and again derive a performance guarantee for decentralized signaling in that setting.  For the correlated setting, we prove that the guarantee's asymptotic dependence upon the number of locations cannot be substantially improved. A comparison of our guarantees for independent locations and for correlated locations reveals the influence of dependence on the performance of decentralized signaling.
}%


\maketitle

\section{Introduction}\label{sec:intro}

In many platforms and service systems, there exists an inherent information asymmetry between the platform operator and its users. Information design, by carefully controlling the revelation of this information, provides an important lever for the system operator to achieve systemwide goals~\citep{lingenbrink2019optimal, alizamir2020warning, candogan2020optimal, kuccukgul2021engineering, anunrojwong2023information, candogan2024value}. Often, in these settings, information available to the operator is multi-dimensional and obtained from various sources. Even if these sources are independent, the operator may face a complex decision about what information to provide so as to best influence subsequent user actions. Moreover, this multi-dimensionality often imposes certain structural constraints on the design of information sharing mechanisms, as we illustrate through the following two vignettes.

\textbf{Vaccination drive:} Consider a county organizing a vaccination drive for its residents, who can visit one of multiple service centers to get vaccinated. Due to factors such as the number of available healthcare professionals at a location, or a sudden increase in demand, the wait times at different service centers are typically different. To guide residents' choice of which service center to visit (and to maximize the likelihood that they visit one), the county considers sharing information about the wait times at different service centers through its website. One approach involves collecting information from different service centers and providing a distilled summary in a centralized fashion. Alternatively, each service center can share information about its own wait times in a {\em decentralized} fashion. Given the data collection and coordination costs associated with the former approach, the county would like to know how effective the latter approach is in maximizing the proportion of vaccinated residents.~\Halmos

\textbf{Seller quality labels:} Online retail platforms like Amazon often use labels such as ``Best seller'' to provide users with information about seller quality, with the goal of maximizing the probability of a sale. The platform could apply such labels strategically based on the qualities of all the sellers being considered by a buyer. However, such an approach to labeling would necessarily be hard to communicate to the buyers. Furthermore, by applying labels on a seller depending on other sellers' quality, the platform's decisions would appear opaque to the sellers. Thus, for the sake of simplicity and transparency, the platform might prefer a labeling approach where labels are applied to each individual seller based solely on her own quality. Again, questions arise as to how to design such a {\em distributed} labeling approach for providing seller quality information, and furthermore, compared to the more complex and opaque approach, how effective this approach is.~\Halmos

As these examples illustrate, the desire for decentralized information sharing mechanisms could arise from the need for structural simplicity, transparency, or due to high coordination and data collection costs. This raises a number of questions. First, what are the powers and limits of decentralized information sharing mechanisms, as compared to more centralized, but complex, ways of information sharing? Relatedly, can one determine conditions under which decentralized mechanisms have provable performance guarantees? Finally, how can we design such decentralized mechanisms with good performance guarantees?

\subsection{Model summary and contributions}

We address the above questions in the framework of Bayesian persuasion~\citep{rayo2010optimal, kamenicaG2011, bergemann2016bayes, bergemann2019information}. Specifically, we consider a model where a service system operator seeks to convince a potential customer to enter the service system, composed of multiple distinct locations. (While we couch our model in the terminology of a service system, our model is applicable to more general settings.) The customer may decide to join the system at any one of the locations or to not join at all. The customer seeks to maximize her own expected utility, which depends on the state of the location that she joins, provided she does indeed join one.  ``State'' could correspond to, for example, how long it would take the customer to be served at a location,  or the quality of service that she would receive there.  The customer may also opt not to join any location, in which case she receives a utility of zero. At the time of her decision, the customer does not know the states of the locations and must rely upon information provided by the system. If the customer joins, the system receives a payoff that depends on the location but not its state. To maximize this payoff, the system decides what information to give to the customer.

We consider two different modes in which information can be provided to the customer. In the {\em centralized} mode, the system operator has access to the states of all the locations and uses that to generate information that is given to the customer.  In the other {\em decentralized} mode, the system operator instructs each individual location to generate information based {\em only} on the location's own state and to directly provide that information to the customer. In this mode, the operator may give instructions to the locations on how to do this, but the actual execution is left to the locations, which act independently of each other. In either mode, the information may be tailored to entice the customer to join the system. However, the customer will not simply take the information at face value, but rather will act strategically in her own interest in response to whatever is provided. 

We wish to understand the relative performance of the two modes. How close is the system's payoff under the decentralized mode to that under the centralized mode? 
It stands to reason that the answer to this question will be affected by whether there is dependence among the individual locations' states. Indeed, it is natural to expect that if locations' states are stochastically dependent, then there will be greater value in coordination across the system. On the other hand, if locations' states are independent, then one might reasonably hope that the decentralized mode will perform almost as well as the centralized mode. So, how does dependence influence what performance guarantees can be made for the decentralized mode relative to the centralized mode?

In this paper, we address the preceding questions.
Our main contributions are as follows:
\begin{enumerate}
    \item \textbf{Optimal mechanisms for independent locations:} For the setting with independent locations, we initially focus our attention on the case where the system's preferences are homogeneous across locations. In particular, the system seeks to maximize the probability that the customer joins the system, without concern for {\em where} the customer joins. Adopting terminology from queueing theory (although we do not model queueing dynamics), we refer to this probability as the ``throughput.'' We formulate and solve two versions of the system's problem of signaling to maximize the throughput: one corresponding to the centralized information sharing mode and one corresponding to the decentralized information sharing mode. Our solution to the centralized signaling problem follows a fairly standard approach that involves applying the revelation principle to obtain a linear programming formulation. On the other hand, the decentralized signaling problem poses a number of challenges. In particular, the revelation principle cannot be directly applied. As our first main contribution, we provide in Theorem~\ref{thm:decentralized_isolated_equivalence} an efficient procedure to compute a decentralized mechanism that maximizes the throughput among all such mechanisms. To obtain this, we first reduce the decentralized signaling problem to a non-linear optimization problem in which the signal space is the set of $K$-dimensional {\em binary} vectors, where $K$ is the number of locations in the system. We then establish that the solution to that non-linear problem can be obtained from the solutions of $K$ distinct signaling problems, where each location is considered in {\em isolation}. In particular, each of these isolated signaling problems can be solved as a simple linear program.

    \item \textbf{Tight performance guarantees for independent locations:} The analysis used to solve the decentralized problem also helps us to compare the throughput of an optimal decentralized signaling mechanism, the optimal throughputs of the $K$ distinct isolated-location systems, and the throughput of an optimal centralized signaling mechanism.  This comparison allows us to prove in Theorem~\ref{thm:cost-of-decentralization} that for any system with $K$ independent locations, the throughput of an optimal decentralized signaling mechanism is at least $1 - (1 - 1/K)^K$ times the throughput of an optimal centralized signaling mechanism. 
    For a system with two locations ($K=2$) the preceding expression is $3/4$, and the expression decreases as the number of locations $K$ increases.  In the limit as the number of locations goes to infinity, the expression approaches $1-e^{-1} \approx 0.63$.  Hence, even if there is a large number of locations, decentralized signaling yields throughput that is at least 63\% of the throughput from centralized signaling. We also construct an example that shows the constant $1 - (1 - 1/K)^K$ cannot be improved. 

    For the setting where the system has heterogeneous preferences across its locations (so that its payoff depends upon where the customer joins), our approach based on solving distinct isolated-location systems does not yield an {\em optimal} decentralized mechanism. Nevertheless, under an additional condition on the customer's utility functions, in Theorem~\ref{thm:cost-dec-heterogeneous} we establish that the decentralized signaling mechanism so-constructed yields the same multiplicative guarantee of $1 - (1 - 1/K)^K$ on the system's payoffs in the heterogeneous setting. Together, these two results (Theorems~\ref{thm:cost-of-decentralization} and~\ref{thm:cost-dec-heterogeneous}) constitute our second main contribution, providing a guarantee that, under independence, the performance of decentralized signaling is not too far from the performance of centralized signaling.

    \item \textbf{Worst-case bounds for dependent locations:} To understand the effect of stochastic dependence on the performance of decentralized signaling, we relax the assumption that the states of locations are independent.  In a setting where the states of locations follow an arbitrary joint distribution, we prove that the throughput of an optimal decentralized signaling mechanism is at least $1/K$ times the throughput of an optimal centralized signaling mechanism for any $K$-location system. 
    As our third main contribution, we show in Theorem~\ref{thm:correlated-upper-bound} that the asymptotic dependence of this guarantee cannot be significantly improved in the worst case. We establish this result by providing an example for which the throughput of an optimal decentralized mechanism is at most $\mathcal{O}(\log K/K)$ times the throughput of an optimal centralized mechanism.  Our multiplicative guarantee of $1/K$ for decentralized signaling in the presence of dependence across locations is markedly weaker than our guarantee of $1 - (1 - 1/K)^K$ for systems with independent locations. This difference is particularly stark when the number of locations $K$ is large. As mentioned above, for $K=2$, optimal decentralized signaling is guaranteed to yield at least 75\% of the centralized optimal throughput when locations are independent.  Without an assumption of independence, this guarantee is instead 50\%. For  
$K=10$, these numbers become about 65\% and 10\%. For    
$K=100$, they become about 63\% and 1\%. 
Theorem~\ref{thm:correlated-upper-bound} shows these guarantees of 50\%, 10\%, and 1\% for systems with dependent states and $K=2$, 10, and 100 locations {\em cannot} be made better than 67\%, 26\%, and 5\% respectively.
This helps to reinforce and quantify the intuition that there is greater benefit to coordination among locations when there is dependence across locations, and that this effect is stronger when there are more locations among which to coordinate.
\end{enumerate}

In sum, our results show that, with independent sources of information, decentralized information sharing mechanisms provide a constant factor guarantee on the performance, regardless of the number of information sources. Furthermore, in such settings, there exist simple and efficient ways to find decentralized mechanisms with good performance guarantees. On the other hand, when the information sources are strongly dependent, there are substantial advantages to centralization, especially when there are many sources.

\subsection{Related work}

Our work belongs to a large literature on using the methodology of Bayesian persuasion and information design~\citep{rayo2010optimal, kamenicaG2011, bergemann2016bayes, bergemann2019information, dughmi2017algorithmic} to study information sharing in practical settings. We refer the reader to the survey by~\citet{dughmi2017survey} for an overview and the one by~\citet{kamenica2019} for a wide range of applications of this approach. 

Given our focus on decentralized signaling in multi-location systems, where each location shares information based on its local state, our work is related to the topic of persuasion with multiple senders. Here, the literature has focused on two types of models. In the first type, multiple senders design ``signals'' revealing information about a common underlying state; in particular, the senders are {\em a priori} informationally identical. Our work, with its focus on limited local information at each location and decentralized signaling, stands in stark contrast to these models. To describe a few papers in this vein, \citet{gentzkowK2017, gentzkow2017bayesian} focus on  competing senders with rich ``Blackwell-connected'' signal spaces, where any single sender can design signals with as much information as any other sender. (\citet{liN2018, liN2021} and \citet{wu2023} study sequential versions of the same competitive setting.) These papers assume the availability of a common randomization or a sunspot variable to arbitrarily correlate the signals revealed by the senders. On the other hand, \citet{hossainWLCPX2024} take a computational perspective on the same model under independent signaling by senders, and show that even computing a sender's best response is NP-hard. 

The second type of models with multiple senders studies persuasion where a number of senders independently reveal information about a personal/local state; our work is more closely related to these papers. As examples, \citet{au2021competitive} and \citet{boleslavksyC2015, boleslavskyC2018} study competitive outcomes in models with two senders, with each sender's state taking binary values.  \citet{au2020competitive} study competition among multiple symmetric senders, with each sender's state being independently and identically distributed taking an arbitrary (but finite) number of values. \citet{duTWZ2024} extend the latter model to asymmetric senders. The focus of all these works is on the competitive setting, where the goal is to identify an equilibrium outcome where each sender designs a signaling mechanism to maximize her own expected payoff, fixing other senders' choice of signaling mechanisms. Our work can be envisioned in this framework as one where (i) all the senders have perfectly-aligned incentives and (ii) the senders can design their respective signaling mechanisms {\em cooperatively} prior to observing their local state. We remark that the optimal decentralized signaling mechanism in our setting also constitutes an equilibrium under this framework.

More broadly, our work studies information design under the structural restriction of decentralization. A number of papers have imposed other structural restrictions on the signaling mechanisms, including public persuasion~\citep{dughmi2017algorithmic, yang2019information, candogan2019}, monotone persuasion~\citep{kolotilin2024monotonepersuasion, mensch2021}, and credible persuasion~\citep{linL2024}. 
Closer to our work, \citet{gradwohlHHS2022} study persuasion in a model similar to ours, under the restriction that the number of signals available to the sender is less than the number of locations (or ``agents'' in their parlance). The authors provide an algorithm for efficiently computing a centralized signaling mechanism with a guarantee akin to our Theorem~\ref{thm:cost-dec-heterogeneous}. This algorithm requires solving a bilevel optimization problem, where in the second level, each agent formulates a linear program to solve for a signaling mechanism based on its own state. These linear programs are similar to the (isolated) linear programs at each location that we solve in the case of independent locations to compute decentralized signaling mechanisms with good guarantees. However, there are a number of differences. First, while \citet{gradwohlHHS2022} allow for state-dependent sender preferences, their focus is on efficient computation of {\em centralized} signaling mechanisms. In contrast, we consider a sender with state-independent preferences, but focus on obtaining decentralized signaling mechanisms. Second, the linear programs in \citep{gradwohlHHS2022} each require as input a parameter obtained by solving a centralized non-linear program in the first level of the bilevel optimization problem. Thus, unlike our setting, these linear programs are not truly {\em independent}. (Our setting can alternatively be thought of setting all the parameter values equal to one; however, this setting of parameter values is infeasible for the problem they consider.) Third, their algorithm combines the signaling mechanisms obtained from solving these linear programs in a complicated manner to obtain a (centralized) signaling mechanism with a specified number of signals. In contrast, we allow each location to independently and directly implement the signaling mechanisms found by solving the linear programs; this, however, necessitates a number of signal vectors that is exponential in the number of locations.  

In a recently posted working paper, \citet{arieli2025decentralized} also tackle the problem of decentralized persuasion. They analyze a model consisting of multiple senders with a common utility who each receive a conditionally independent signal about a common underlying state. The authors show that, barring instances where a single partially informed sender can achieve the same utility as one with full state information, decentralized mechanisms cannot obtain the same utility as centralized mechanisms (using our terminology). Furthermore, for the case of binary states, the authors show that, as the number of senders grows sufficiently large, the optimal decentralized signaling mechanism obtains a constant fraction of the optimal centralized payoff. Our model with independent locations can be fit into their general framework, but would require an exponential (in the number of senders/locations) number of states; as mentioned earlier, they provide constant factor guarantees only for the case of two states. More importantly, the constant factor guarantee they obtain for the case of binary states is instance-dependent (depending on the prior), whereas our constant factor guarantees for the case of independent locations are instance-independent with a $1-e^{-1}$ lower bound. Additionally, we provide a tight guarantee for the case of correlated locations.

Finally, our work contributes to a growing literature analyzing the powers and limits of decentralization on  the design of markets and platforms~\citep{aouadSY2023, kapoorPSM2022, cachon2023pricing, filippasJS2023, atasu2024price}.

The remainder of the paper is organized as follows. Section~\ref{sec:system} presents our model. Section~\ref{sec:opt-dec} introduces decentralized signaling mechanisms and shows how to obtain an optimal decentralized signaling mechanism. Section~\ref{sec:cost-dec} contains our results on the performance of decentralized signaling  for systems with independent locations.
Section~\ref{sec:heterogeneous} extends our analysis to settings where the system operator has heterogeneous preferences across locations. 
Section~\ref{sec:correlated} analyzes systems where states of locations are dependent.
Section~\ref{sec:proofs} contains proofs and supporting results.

\section{The model} \label{sec:system}

Consider a system comprising multiple service locations, one of which
an arriving customer may visit. We let $\comp = \{1, \dots, K\}$ be
the set of locations. Each location $k \in \comp$ has a random state
$\hat{\omega}_k$ that takes values in a finite set $\Omega_k$. The
random state of the system
$\hat{\omega} = (\hat{\omega}_1,\dots,\hat{\omega}_K)$ takes values in
the set $\Omega = \Omega_1 \times \dots \times \Omega_K$. Let $\mu$ be
the joint distribution of the system state; that is,
$\mu(\omega) = \prob(\hat{\omega} = \omega)$ for
$\omega = (\omega_1, \dots, \omega_K) \in \Omega$. Except in
Section~\ref{sec:correlated}, we assume that the states of the
locations are independent, so
$\mu(\omega) = \prod_{k\in \mathcal{K}} \mu_k(\omega_k)$, where
$\mu_k$ denotes the distribution of the state of location~$k$; that
is, $\mu_k(\omega_k) = \prob(\hat{\omega}_k = \omega_k)$ for
$\omega_k \in \Omega_k$.

Upon arrival, the customer decides either to obtain service at (i.e., ``join'') one of the locations or to leave the system without obtaining service. Let $\comp_0 \coloneq \comp \cup \{0\}$ denote the set of actions available to the customer, where the action $a = 0$ implies that the customer leaves the system without obtaining service, and the action $a = k \in \comp$ implies that the customer joins location $k$. We normalize the customer's utility for leaving without obtaining service to zero and assume that the customer's utility from joining a location $k \in \comp$ depends on the state of that location, but not on the states of other locations. Formally, the customer's utility $u(\omega, a)$ from taking the action $a \in \comp_0$ at system state $\omega = (\omega_1, \dots, \omega_K)$ is given by
\begin{align*}
 u(\omega, a) = \sum_{k \in \comp} \ind\{a = k\} h_k(\omega_k).
\end{align*}
for some functions $h_k\colon \Omega_k \to \reals $.  Note $u(\omega, 0)=0$. We summarize the preceding service system using the notation $\sys[(\mu, h_1, \dots, h_K)]$, and if the context makes it clear, we say $\sys$ for brevity.

We assume the realized value of the random system state is known to the system, but is not known to the customer when the customer must choose her action.
To induce the customer to join one of the locations, the service system shares information about the system state using a {\em signaling mechanism}. We begin by describing signaling mechanisms in full generality; later we will impose further restrictions on the structure of the signaling mechanisms when focusing on decentralized modes of sharing information in Section~\ref{sec:opt-dec}. A signaling mechanism $(\sign, \sigma)$ consists of a finite set of signals $\sign$ and a function $\sigma\colon\Omega \times \sign \to [0,1]$, where $\sigma(s|\omega)$ denotes the probability with which the system sends signal $s \in \sign$ to the customer when the system state is $\hat{\omega}=\omega \in \Omega$. Let $\mathcal{C}(\sign) \coloneq \{ \sigma \colon \Omega \times \sign \to [0,1] :  \sum_{s \in \sign} \sigma(s|\omega) = 1 \text{ for all $\omega \in \Omega$}\}$. When the context is clear, we abuse terminology and refer to the function $\sigma$ as the signaling mechanism. The finiteness assumption on the set $S$ is without loss of generality: for the centralized setting, it follows from standard results~\citep{kamenicaG2011,bergemann2016bayes} that $\min\{|\Omega|, K+1\}$ signals suffice. For the decentralized setting, we argue similarly in Section~\ref{sec:opt-dec} that $|\Omega_k|$ signals suffice for each location $k$.

As is standard in the literature on information design~\citep{kamenicaG2011,bergemann2016bayes}, we assume that the system has the power to {\em commit} to sharing signals using a prespecified signaling mechanism $\sigma \in \mathcal{C}(\sign)$. Let $\hat{s}$ be the random $\sign$-valued signal sent to the customer under $\sigma$, and let $\prob^\sigma$ denote the joint distribution of $(\hat{\omega}, \hat{s})$ under the signaling mechanism $\sigma$, with $\expec^\sigma$ denoting the corresponding expectation.
The customer is Bayesian with prior belief given by $\mu$. After receiving the signal $\hat{s} = s$ from the system, and with the knowledge of $\sigma$, the customer updates her belief and chooses an action $a \in \comp_0$ that maximizes her expected utility $\expec^\sigma[u(\hat{\omega}, a) | \hat{s} = s]$ with respect to her posterior belief.

Formally, given a finite set of signals $\sign$, a customer's strategy is specified as $f\colon\comp_0 \times \sign \to [0,1]$, where $f(a|s)$ denotes the probability with which the customer chooses action $a \in \comp_0$ after seeing the signal $s \in \sign$. We let $\mathcal{F}(\sign) \coloneq \{ f\colon\comp_0 \times \sign \to [0,1] : \sum_{a \in \comp_0} f(a|s)  = 1 \text{ for all } s \in \sign\}$ denote the set of customer strategies under the set $\sign$ of signals. 

Given a signaling mechanism $\sigma \in \mathcal{C}(\sign)$, a strategy $f \in \mathcal{F}(\sign)$ is optimal under $\sigma$ for the customer if, for any signal 
$s \in \sign$, the strategy puts positive weight only on those actions that maximize the customer's expected utility (with respect to the posterior belief). That is, $f \in \mathcal{F}(\sign)$ is optimal under $\sigma$ for the customer   if for all $s \in \sign$ with $\prob^\sigma(\hat{s} = s) > 0$  and $a \in \comp_0$,
\begin{align}\label{eq:optimal-strategy}
    f(a|s) > 0 \implies a \in \arg\max_{a' \in \comp_0} \expec^\sigma[u(\hat{\omega},a') | \hat{s} = s].
\end{align}
When the context is clear, we will sometimes simply say that $f$ is optimal.

Throughout, we frequently make use of the fact that for $k \in \comp$ we have
\begin{align*} \expec^\sigma[u(\hat{\omega},k) | \hat{s} = s]  & = \expec^\sigma [h_k (\hat{\omega}_k) | \hat{s} = s] \\
 & = \frac{1}{\prob^\sigma(\hat{s} = s)}
\expec^\sigma [ h_k(\hat{\omega}_k) \ind \{ \hat{s} = s \} ] \\
 & = \frac{1}{\prob^\sigma(\hat{s} = s)}
\sum_{\omega \in \Omega} \mu(\omega) \sigma(s|\omega) h_k(\omega_k),
\end{align*} 
provided that $\prob^\sigma(\hat{s} = s) = \sum_{\omega \in \Omega} \mu(\omega) \sigma(s|\omega) >0$.

The system seeks a signaling mechanism that maximizes the {\em throughput}, i.e., the probability that the customer joins one of the locations, assuming the customer responds with an optimal strategy. (In Section~\ref{sec:heterogeneous}, we generalize this to allow the system's preferences to be heterogeneous across locations so that the system is concerned not only with whether or not a customer joins one of its locations, but also with which location the customer joins.) To elaborate, under a signaling mechanism $\sigma \in \mathcal{C}(\sign)$ and a customer strategy $f \in \mathcal{F}(\sign)$, let $\hat{a}$ denote the customer's action. The joint distribution $\prob^{\sigma,f}$ of $(\hat{\omega}, \hat{s}, \hat{a})$ depends upon the system's choice of $\sigma$ and the customer's choice of $f$ (and also the state distribution $\mu$, which is fixed throughout). The throughput is defined as
\begin{align}\label{eq:thput}
    T(\sigma, f) &\coloneq \expec^{\sigma, f} \left[ \sum_{a \in \comp} \ind\{ \hat{a} = a\}\right]\notag \\
   &=\sum_{a \in \comp} \prob^{\sigma, f}\left( \hat{a} = a\right) \notag\\
    &= \sum_{\omega \in \Omega} \sum_{s \in \sign} \sum_{a \in \comp} \mu(\omega) \sigma(s|\omega) f(a|s).
\end{align}
When the system uses mechanism $\sigma$ and the customer uses strategy $f$, the term $\mu(\omega)\sigma(s|\omega) f(a|s)$ above represents the probability that the state is $\omega$, the signal is $s$, and the customer action is $a$; that is, the probability that $(\hat{\omega}, \hat{s}, \hat{a}) = (\omega,s,a)$. 

The system's goal is to choose a signaling mechanism $\sigma \in \mathcal{C}(\sign)$ that maximizes the throughput $T(\sigma, f)$, where $f \in \mathcal{F}(\sign)$ is an optimal strategy satisfying~\eqref{eq:optimal-strategy} with ties broken in favor of the system. This assumption on tie-breaking is common in the literature, and avoids having to resort to $\epsilon$-optimal mechanisms to ensure strict preference in generic settings. Formally, we can write the system's problem as
\begin{align}\label{eq:system-centralized-problem}
    \thput = \max_{\sign} \max_{\sigma \in \mathcal{C}(\sign), f \in \mathcal{F}(\sign)} \quad  T(\sigma, f) \quad \text{s.t.\  $f$ satisfies \eqref{eq:optimal-strategy} under $\sigma$}.
\end{align}
While~\eqref{eq:system-centralized-problem} describes the system's decision problem in complete generality, the formulation can be considerably simplified. By a standard revelation principle argument~\citep{bergemann2016bayes}, it suffices to consider (i) {\em direct} signaling mechanisms $\sigma$ with $\sign = \comp_0$, i.e., for which signals are action recommendations, and (ii) for which the {\em obedient} strategy $f_{ob} \in \mathcal{F}(\comp_0)$, defined as $f_{ob}(k|k) = 1$ for all $k \in \comp_0$, is optimal, i.e., it is optimal for the customer to follow the action recommended by $\sigma$. Focusing only on such mechanisms and the obedient strategy, we can reformulate the system's decision problem as a linear program using an approach described by, e.g., \citet{bergemann2019information}.  This is detailed in the following proposition.
\begin{proposition}
The problem~\eqref{eq:system-centralized-problem} can be equivalently posed as the following linear program:
\begin{align}
    \max_{\sigma\in \mathcal{C}(\comp_0)}   \sum_{\omega \in \Omega} \sum_{k \in \comp} &\ \mu(\omega) \sigma(k|\omega) \notag \\
\textup{s.t. }   \sum_{\omega \in \Omega} \mu(\omega) \sigma(k|\omega) h_k(\omega_k)  &\geq \sum_{\omega \in \Omega}  \mu(\omega) \sigma(k|\omega) h_{\ell}(\omega_{\ell})  \text{ for all $k, \ell \in \comp$} \notag \\
\sum_{\omega \in \Omega} \mu(\omega) \sigma(k|\omega) h_k(\omega_k)  &\geq 0  \text{ for all $k \in \comp$} \notag \\
\sum_{\omega \in \Omega}  \mu(\omega) \sigma(0|\omega) h_k(\omega_k) &\leq 0 \text{ for all $k \in \comp$.}\label{eq:centralized-lp}
\end{align}
\end{proposition}
\proof{Proof.} The revelation principle argument states that it suffices to consider signaling mechanisms $\sigma \in \mathcal{C}(\comp_0)$ for which the obedient strategy $f_{ob}$ is optimal, i.e., $f_{ob}$ satisfies~\eqref{eq:optimal-strategy}. For any such $\sigma$, this condition holds  if and only if $\expec^\sigma[ u(\hat{\omega},k) | \hat{s}= k] \geq \expec^\sigma[ u(\hat{\omega},\ell) | \hat{s}= k]$ for all $\ell,k \in \comp_0$ with $\prob^\sigma(\hat{s} = k) > 0$. These latter conditions are equivalent to requiring, for all $k \in \comp$,
    \begin{align*}
        \expec^\sigma\left[ h_k(\hat{\omega}_k)\ind\{\hat{s} = k\}\right] &\geq \expec^\sigma\left[ h_{\ell}(\hat{\omega}_{\ell})\ind\{\hat{s} = k\}\right] \quad \text{for all $\ell \in \comp$},\\
        \expec^\sigma\left[ h_k(\hat{\omega}_k)\ind\{\hat{s} = k\}\right] &\geq 0, \\
        \expec^\sigma\left[ h_k(\hat{\omega}_k)\ind\{\hat{s} = 0\}\right] &\leq 0,
    \end{align*}
    where we have used the fact that, for all $k \in \comp_0$, $\expec^\sigma[u(\hat{\omega},0) \ind\{ \hat{s}=k\}] = 0$ and $\expec^\sigma[u(\hat{\omega},\ell) \ind\{ \hat{s}=k \}] = \expec^\sigma[h_{\ell}(\hat{\omega}_{\ell}) \ind\{ \hat{s}=k \}]$ for $\ell \in \comp$. 
    Since  $\expec^\sigma\left[ h_{\ell}(\hat{\omega}_{\ell})\ind\{\hat{s} = k\}\right] = \sum_{\omega \in \Omega} \mu(\omega) \sigma(k|\omega) h_{\ell}(\omega_{\ell})$ for all $\ell \in \comp$ and $k \in \comp_0$, we obtain the constraints in the proposition statement. Furthermore, the throughput under $\sigma$ and the obedient strategy $f_{ob}$ is given by 
    \begin{align*}
        T(\sigma, f_{ob}) = \expec^\sigma\left[ \sum_{k \in \comp} \ind\{\hat{s} = k\}\right] = \sum_{\omega \in \Omega} \sum_{k\in \comp} \mu(\omega) \sigma(k|\omega).
    \end{align*}
    This completes the proof.
\Halmos\endproof
In general, implementation of a signaling mechanism as detailed above requires access to the entire system state. The centralized nature of such a mechanism necessitates coordination and communication among the locations, and may present challenges to practical implementation. To address this, we may impose the requirement of {\em decentralization} as described next in Section~\ref{sec:opt-dec}. 

\section{Decentralized signaling}\label{sec:opt-dec}

In this section we introduce decentralized signaling and show how to obtain an optimal decentralized signaling mechanism.
Under decentralized signaling, we require each location to independently and directly share information about its state with the customer. Formally, we say a signaling mechanism $(\sign, \sigma)$ is decentralized if it has the following structure: (i) the set $\sign$ can be written as $\sign = \sign_1 \times \dots \times \sign_K$, where $\sign_k$ is interpreted as the set of signals sent by location $k$; and (ii) for any $s = (s_1, \dots, s_K) \in \sign$ and $\omega = (\omega_1, \dots, \omega_K) \in \Omega$, we have $\sigma(s|\omega) = \prod_{k \in \comp} \sigma_k(s_k| \omega_k)$, where $\sum_{s_k \in \sign_k} \sigma_k(s_k|\omega_k) = 1$ and $\sigma_k(s_k|\omega_k) \geq 0$.
In other words, a location $k \in \comp$ with state $\omega_k \in \Omega_k$ sends signal $s_k \in \sign_k$ with probability $\sigma_k(s_k|\omega_k)$, independent of the states and signals of other locations. For any $\sign = \sign_1 \times \dots \times \sign_K$, we define $\mathcal{D}(\sign) \subseteq \mathcal{C}(\sign)$ as the set of all $\sigma \in \mathcal{C}(\sign)$ that are decentralized: $\mathcal{D}(\sign) \coloneq \{ \sigma \in \mathcal{C}(\sign) : \sigma(s|\omega) = \prod_{k \in \comp} \sigma_k(s_k| \omega_k) \text{ for $s \in \sign$ and $\omega \in \Omega$}\}$. Note that the set of decentralized signaling mechanisms is a subset of the set of centralized signaling mechanisms.

The central goal of this paper is to understand the performance of decentralized signaling relative to that of centralized signaling.  As a first attempt to address this objective, we may ask if there are simple decentralized signaling mechanisms that guarantee good performance.  One simple decentralized mechanism is the {\em full-information mechanism}, in which each location truthfully reports its own state to the customer.  Another simple decentralized mechanism is the {\em no-information mechanism}, in which each location provides no information to the customer. (Providing no information may be viewed as a decentralized signaling mechanism by taking each location's signal space to be a singleton.)
The next example shows that the throughput of these two mechanisms may be 
arbitrarily bad relative to that of the best centralized signaling mechanism.  Hence, to have any hope of guaranteed good performance from decentralized signaling, we need to find decentralized mechanisms that are better than the full-information and no-information mechanisms.  This provides motivation to find the best decentralized signaling mechanism, which we take up after the example.

\begin{example}[Inadequacy of full-information and no-information mechanisms]\label{ex:inadequacy}
    Consider a $K$-location system, where each location $k$ has state space $\Omega_k = \{0, 1\}$ with $\mu_k(1) = \prob(\hat{\omega}_k = 1) = p$. Assume $h_k(1) = X > 1$ and $h_k(0) = -1$. Recall that we assume that location states are independent. 
    
    Under the full-information mechanism, the customer will join some location whenever there is at least one location in state $1$, and hence the full-information throughput is  
    $
        \thput_{\mathsf{full}} = 1 - (1-p)^K \eqcolon q.
    $
    Now, consider the centralized mechanism $\sigma$ defined as follows: Let $\hat{\compn} \coloneq \{ k \in \comp : \hat{\omega}_k = 1\}$ denote the set of all locations in state $1$. If $\hat{\compn}  \neq \emptyset$, then $\sigma$ selects a location uniformly at random from the set $\hat{\compn}$ and asks the customer to join it. On the other hand, if $\hat{\compn}  = \emptyset$, then $\sigma$ selects a location uniformly at random from $\comp$, and asks the customer to join it. For $\sigma$ to satisfy the obedience conditions, we require $q X + (1- q) (-1) \geq 0$ or, equivalently, $q \geq (X+1)^{-1}$. For $p = p^* \coloneq 1 - \left(\frac{X}{X+1}\right)^{1/K}$, we obtain $q^* = 1 - (1- p^*)^K = (X+1)^{-1}$. So, for $p=p^*$ the throughput of $\sigma$ is $1$, and hence $\thput = 1$. We conclude that for $p = p^*$, we have 
    $\thput_{\mathsf{full}} / \thput = q^*/1 = (X+1)^{-1}$.
    Therefore, the throughput of the full-information mechanism can be arbitrarily bad (by taking $X$ large enough), even for $K=2$. 
    
    With $p = p^*$, it is straightforward to check that under the no-information mechanism, the customer's expected utility from joining any location is strictly negative; i.e., $p^* X + (1-p^*) (-1)<0$.  This implies that the no-information throughput is zero; $\thput_{\mathsf{no}} = 0$ and $\thput_{\mathsf{no}}/\thput = 0$. 
    To summarize, this example shows that both the full-information mechanism and the no-information mechanism can perform arbitrarily poorly relative to the best centralized signaling mechanism.%
\Halmos\end{example}

The problem of finding an optimal decentralized signaling mechanism can be formulated as follows:
\begin{align}
\label{eq:system-decentralized-problem}
\thputD =  \max_{\sign = \sign_1 \times \dots \times \sign_K} \max_{\sigma \in \mathcal{D}(\sign), f \in \mathcal{F}(\sign)} \quad  T(\sigma, f)\quad \text{s.t.\ $f$ satisfies \eqref{eq:optimal-strategy} under $\sigma$}.
\end{align}
It is immediate from \eqref{eq:system-centralized-problem} and \eqref{eq:system-decentralized-problem} that $\thputD \le \thput$. Below we focus on analyzing and solving \eqref{eq:system-decentralized-problem} and some of its variants, and answering the question: How close is $\thputD$ to $\thput$?

Before proceeding, we provide a roadmap of our approach to~\eqref{eq:system-decentralized-problem}.  We will first argue that to solve~\eqref{eq:system-decentralized-problem}
it suffices to consider only decentralized signaling mechanisms in which the signal space for each location is finite.
This yields formulation \eqref{eq:decentralized-nlp} below.  We will subsequently show in Lemma~\ref{lem:decentralized_signal_space} that we may further restrict attention to decentralized mechanisms in which the signal space for each location is $\{0,1\}$ and to customer strategies of a particular form, which we will describe later. From this, we will obtain formulation~\eqref{eq:sys_org_obj}.  We then will use Lemmas~\ref{lem:thput-ind-fd} and~\ref{lem:decentralized-obedience} to provide an expression for the objective function of~\eqref{eq:sys_org_obj} and to characterize the feasible region of~\eqref{eq:sys_org_obj}.   Finally, we will prove in Theorem~\ref{thm:decentralized_isolated_equivalence} that we can solve~\eqref{eq:sys_org_obj} --- and hence~\eqref{eq:system-decentralized-problem} --- by solving $K$ simple linear programs.

Analyzing \eqref{eq:system-decentralized-problem} is challenging 
because the revelation principle fails to hold for decentralized mechanisms. To see this, observe that the revelation principle states that for any signaling mechanism $\sigma \in \mathcal{C}(\sign)$ and optimal strategy $f \in \mathcal{F}(\sign)$, there exists a direct signaling mechanism $\sigma' \in \mathcal{C}(\comp_0)$ for which $f_{ob}$ is optimal and obtains the same throughput. However, there is no guarantee that, for a decentralized mechanism $\sigma \in \mathcal{D}(\sign)$, the equivalent mechanism $\sigma'$ so obtained is decentralized. 

For the setting of centralized mechanisms, the revelation principle provides two main advantages: (i) a canonical set of signals, equal to the set of actions, and (ii) a direct link of each value of a signal to an action for the customer (through obedience). Without the revelation principle in the decentralized setting, we cannot rely on either of these aspects to simplify~\eqref{eq:system-decentralized-problem}.

Nevertheless, we argue next that it suffices to consider decentralized signaling mechanisms for which $|\sign| = |\Omega|$, with $|\sign_k| = |\Omega_k|$ for each $k \in \comp$. We do this by showing that, starting with any decentralized signaling mechanism $\sigma$, we can obtain another decentralized signaling mechanism $\psi$, using at most $|\Omega_k|$ signals for each location $k$, with at least as much throughput as in $\sigma$. To this end, fix a decentralized signaling mechanism $\sigma$, and consider a location $k$.  Let $\mu_{-k}(\omega_{-k}) = \prod_{\ell \neq k} \mu_{\ell}(\omega_{\ell})$ and $\sigma_{-k} = (\sigma_{\ell} : \ell \neq k)$. For any distribution $\eta_k$ over $\omega_k \in \Omega_k$, assuming (i) location $k$ does not share any information, (ii) the other locations share signals according to $\sigma_{-k}$ and (iii) all other locations' states are distributed as $\mu_{-k}$, the total throughput can be written as a function $T(\eta_k | \mu_{-k}, \sigma_{-k})$. Suppose now that $\mu_k$ is the distribution of $\omega_k$ and that location~$k$ can share information with the customer by selecting a mechanism $\psi_k(s_k|\omega_k)$, but (ii) and (iii) still hold. Results of \citet{kamenicaG2011} state that the optimal throughput for the system (where the optimization is over $\psi_k$) is given by $\overline{T}(\mu_k | \mu_{-k}, \sigma_{-k})$, where $\overline{T}(\cdot | \mu_{-k}, \sigma_{-k})$ is the concavification of $T(\cdot | \mu_{-k}, \sigma_{-k})$. Furthermore, Proposition~4 in their Supplemental Appendix implies that this throughput can be attained by a signaling mechanism $\psi_k$ using at most $|\Omega_k|$ signals. Thus, starting with any decentralized signaling mechanism $\sigma$, we obtain a signaling mechanism $\sigma'$, where all locations $\ell \neq k$ use the mechanism $\sigma_{\ell}$ and location $k$ uses the mechanism $\psi_k$ (with at most $|\Omega_k|$ signals), and which has at least as much throughput as $\sigma$. Repeating the above steps for all other locations sequentially, we conclude that there exists a signaling mechanism $\psi$ with at least as much throughput as $\sigma$, and where each location $k$ uses at most $|\Omega_k|$ signals. So, without loss of optimality, we can assume that $\sign_k = \fsign_k := 
\{ 0, 1, \dots, |\Omega_k | - 1 \}$ for each $k \in \comp$ and $\sign = \fsign := \fsign_1 \times \dots \times \fsign_K$. 

Based on this argument, the problem~\eqref{eq:system-decentralized-problem} can be formulated as follows:
\begin{align}
\label{eq:decentralized-nlp}
\thputD =  \max_{\sigma \in \mathcal{D}(\fsign), f \in \mathcal{F}(\fsign)} \quad  T(\sigma, f)\quad \text{s.t.\ $f$ satisfies \eqref{eq:optimal-strategy} under $\sigma$.}
\end{align}
It is instructive to write the constraints in \eqref{eq:decentralized-nlp} explicitly in terms of the entries of $\sigma$ and $f$. Doing so yields the following equivalent formulation:
\begin{align}
    \thputD = &\max_{\sigma \in \mathcal{D}(\fsign), f \in \mathcal{F}(\fsign)}  
    \;\; T(\sigma,f) \notag \\
    \text{s.t. } & \sum_{\omega \in \Omega} \mu(\omega) \prod_{k \in \comp} \sigma_k(s_k | \omega_k) \sum_{\ell \in \comp} f(\ell|s) h_{\ell}(\omega_{\ell})\notag\\
    &\quad \geq 
    \sum_{\omega \in \Omega} \mu(\omega) \prod_{k \in \comp} \sigma_k(s_k | \omega_k) h_{m}(\omega_{m}) \quad \text{for all $m \in \comp$, $s \in \fsign$}\notag\\
    & \sum_{\omega \in \Omega} \mu(\omega) \prod_{k \in \comp} \sigma_k(s_k | \omega_k) \sum_{\ell \in \comp} f(\ell|s) h_{\ell}(\omega_{\ell}) \geq 0 \quad \text{for all $s \in \fsign$}.
    \tag{\ref{eq:decentralized-nlp}$'$} 
    \label{eq:decentralized-nlp-prime}
\end{align}
Here, the first inequality corresponds to the requirement that, upon receiving any signal vector $s \in \fsign$ from $\sigma$, the customer's expected utility under $f$ should be at least as high as that obtained from joining a location $m \in \comp$. Similarly, the second constraint states that this customer's expected utility should be at least as high as that obtained from not joining any location, i.e., zero. Together, these constraints ensure the optimality of $f$ under $\sigma$.    
The formulation \eqref{eq:decentralized-nlp-prime} allows us to see that there are finitely many constraints in \eqref{eq:decentralized-nlp} because $\fsign$ is finite. Note also that the argument above provides us with a canonical set of signals $\fsign_k$ for each location~$k$. However, unlike the revelation principle, it does not link each signal to an action for the customer. Consequently, \eqref{eq:decentralized-nlp} involves optimizing over the customer's strategy $f$, unlike~\eqref{eq:centralized-lp} for the centralized setting. A second point of difference is that~\eqref{eq:decentralized-nlp} is highly non-linear. Together, these differences make it difficult to obtain an optimal decentralized signaling mechanism by solving problem~\eqref{eq:decentralized-nlp} as posed above.

To address this, we next show that it is sufficient to restrict attention to decentralized signaling mechanisms with a binary vector of signals. This will ultimately allow us to obtain an optimal decentralized signaling mechanism by solving $K$ linear programs. 

\subsection{Reduction to binary signals} \label{subsec:reduc-dec}
In preparation for our next result, let $\uset\coloneq\{0,1\}^K$.  
We denote by $\zero$ and $\one$ the vectors in $\uset$ with all entries equal to $0$ and $1$ respectively. Define $\mathcal{F}_d$ as the set of strategies $f \in \mathcal{F}(\uset)$ satisfying the following two conditions: (i) for all $u \in \uset$ with $u \neq \zero$, we have $f(0|u) = 0$, and (ii) for each $k \in \comp$ and for all $u \in \uset$ with $u_k = 0$, we have $f(k|u) = 0$. 
To understand this definition, note that if the signal space is $\uset$ and a customer uses a strategy in $\mathcal{F}_d$, then upon receiving a signal of (say) $u$, the customer will join some location~$k$ for which $u_k=1$, provided there is at least one such location.  Note also that $\sum_{k \in \comp} f(k|u) = \ind\{u \neq \zero\}$ for $f \in \mathcal{F}_d$.

The following result shows that to analyze the decentralized persuasion  problem~\eqref{eq:decentralized-nlp}, it suffices to consider decentralized signaling mechanisms in $\mathcal{D}(\uset)$ for which a strategy in $\mathcal{F}_d$ is optimal.

\begin{lemma}\label{lem:decentralized_signal_space}
Let $\sigma \in \mathcal{D}(\fsign)$ be a decentralized signaling mechanism with signal space $\fsign = \fsign_1 \times \dots \times \fsign_K$. Then, there exist a decentralized signaling mechanism $\psi \in \mathcal{D}(\uset)$ and a strategy $\hat{f} \in \mathcal{F}_{d}$ such that
 \begin{enumerate} 
 \item $\hat{f}$ is optimal under the signaling mechanism $\psi$; 
 \item $T( \psi,\hat{f}) \geq T(\sigma,f )$ for any strategy $f \in \mathcal{F}(\fsign)$ that is optimal under $\sigma$.
 \end{enumerate}
 \end{lemma}
A complete proof of the lemma can be found in Section~\ref{sec:proofs}. Here, we confine ourselves to providing an intuitive explanation.  Given a decentralized mechanism $\sigma \in \mathcal{D}(\fsign)$, the decentralized mechanism $\psi \in \mathcal{D}(\uset)$ in the lemma works as follows. If, under $\sigma$, location~$k$ would send a signal $s_k \in \sign_k$ such that the customer's conditional expected utility from joining location~$k$ upon receiving signal $s_k$ is non-negative, then location~$k$ sends a signal of $u_k=1$ under $\psi$.   Otherwise, location~$k$ sends a signal of $u_k=0$ under $\psi$.  The proof of the lemma involves checking that, under $\psi$, it is in the best interest of the customer to join some component that sends her a signal of 1 and not to join any component that sends her a signal of 0, and that this yields a throughput no smaller than that of the original $\sigma$.

Using Lemma~\ref{lem:decentralized_signal_space}, we can take $\sign=\uset$ and re-formulate the system's problem of maximizing throughput over decentralized signaling mechanisms as:
\begin{align}
\max_{\sigma\in \mathcal{D}( \uset ),f \in \mathcal{F}_d}~~ T(\sigma,f) \quad \text{s.t.\ $f$ satisfies \eqref{eq:optimal-strategy} under $\sigma$.} \label{eq:sys_org_obj}
\end{align}

\subsection{Eliminating dependence on customer's strategy}\label{sec:eliminate-dependence}
While Lemma~\ref{lem:decentralized_signal_space} provides a simpler set of signals, the preceding optimization problem still involves optimizing over the customer's strategy $f$. To simplify further, our next two results show that the dependence on $f$ can be completely eliminated. In particular, the next lemma provides an expression for the objective function in \eqref{eq:sys_org_obj} that is independent of the customer's strategy $f \in \mathcal{F}_d$.
\begin{lemma}\label{lem:thput-ind-fd} For any $\sigma \in \mathcal{D}(\uset)$ and for any $f \in \mathcal{F}_d$, the throughput $T(\sigma, f)$ defined in \eqref{eq:thput} satisfies:
\begin{align*}
    T(\sigma, f) &= \sum_{u \in \uset\setminus\{\zero\}} \prod_{k \in \comp} \left( \sum_{\omega_k \in \Omega_k} \mu_k(\omega_k) \sigma_k(u_k| \omega_k)\right) \\
    &= 1 - \prod_{k\in \comp} \left( 1 - \sum_{\omega_k \in \Omega_k} \mu_k(\omega_k) \sigma_k(1| \omega_k)\right).
\end{align*}
In particular, for any $\sigma \in \mathcal{D}(\uset)$, the throughput $T(\sigma, f)$ is a constant as a function over $f \in \mathcal{F}_d$. 
\end{lemma}
\proof{Proof.} Fix $\sigma \in \mathcal{D}(\uset)$. For any $f \in \mathcal{F}_d$, we have 
\begin{align*}
    T(\sigma, f) &= \sum_{\omega \in \Omega } \sum_{u \in \uset} \mu(\omega)  \sigma(u|\omega) \sum_{k \in \comp} f(k|u)\\
    &= \sum_{\omega \in \Omega} \sum_{u \in \uset\setminus\{\zero\}}\mu(\omega) \sigma(u|\omega)\\
    &= \sum_{u \in \uset\setminus\{\zero\}} \prod_{k \in \comp} \left( \sum_{\omega_k \in \Omega_k} \mu_k(\omega_k) \sigma_k(u_k| \omega_k)\right).
\end{align*}
Here, the second equality follows from the fact that $\sum_{k \in \comp} f(k|u) = \ind\{u \neq \zero\}$ for all $f \in \mathcal{F}_d$. In the third equality, we have used the fact that $\mu(\omega)\sigma(u|\omega) = \prod_{k \in \comp} \mu_k(\omega_k)\sigma_k(u_k|\omega_k)$  for the decentralized mechanism $\sigma$. This establishes that $T(\sigma, f)$ is a constant over $f \in \mathcal{F}_d$.

Finally, the second expression for $T(\sigma,f)$ in the lemma statement follows because 
\begin{align*} 
\sum_{u \in \uset\setminus\{\zero\}} \prod_{k \in \comp} \sum_{\omega_k \in \Omega_k} \mu_k(\omega_k) \sigma_k(u_k|\omega_k) = 1 - \prod_{k \in \comp} \sum_{\omega_k \in \Omega_k} \mu_k(\omega_k) \sigma_k(0|\omega_k)
\end{align*}
and $\sigma_k(0|\omega_k) = 1 - \sigma_k(1|\omega_k)$ for all $\omega_k \in \Omega_k$ and $k \in \comp$. 
\Halmos\endproof

The preceding lemma implies that to find an optimal decentralized mechanism, it is sufficient to maximize the expressions for $T(\sigma,f)$ in Lemma~\ref{lem:thput-ind-fd} over $\sigma \in \mathcal{D}(\uset)$ for which there exists an $f \in \mathcal{F}_d$ satisfying~\eqref{eq:optimal-strategy} under $\sigma$. The following result exactly characterizes all such mechanisms.
\begin{lemma}\label{lem:decentralized-obedience}
    For any $\sigma \in \mathcal{D}(\uset)$, there exists an $f \in \mathcal{F}_d$ that is optimal under $\sigma$ if and only if either one of the following two conditions holds:
    \begin{enumerate}
        \item[\textup{(I)}] there exists a location $k \in \comp$ with (a) $\mu_k(\omega_k) \sigma_k(0|\omega_k) = 0$ for all $\omega_k \in \Omega_k$; (b) $\sum_{\omega_k \in \Omega_k} \mu_k(\omega_k) h_k(\omega_k) \geq 0$, and (c) for all $\ell \neq k \in \comp$,
        \begin{align*}
        \left(\sum_{\omega_{\ell} \in \Omega_\ell} \mu_\ell(\omega_\ell) \sigma_\ell(0|\omega_\ell)\right) \cdot \sum_{\omega_k \in \Omega_k} \mu_k(\omega_k) h_k(\omega_k) \geq \sum_{\omega_{\ell} \in \Omega_\ell} \mu_\ell(\omega_\ell) \sigma_\ell(0|\omega_\ell) h_\ell(\omega_\ell).
        \end{align*}        
        \item[\textup{(II)}] for all $k \in \comp$, we have
    \begin{align}
    \sum_{\omega_k \in \Omega_k} {\mu_k(\omega_k) {\sigma}_k(1|\omega_k)h_k(\omega_k)} &\geq 0,\label{eq:sys_mod_con1}\\
    \sum_{\omega_k \in \Omega_k} \mu_k(\omega_k) {\sigma}_k(0|\omega_k) h_k(\omega_k) &\leq 0.\label{eq:sys_mod_con2}
\end{align}
\end{enumerate}
\end{lemma}
To interpret these conditions, suppose a signaling mechanism $\sigma \in \mathcal{D}(\uset)$ is such that there exists an optimal $f \in \mathcal{F}_d$. This could arise because there could exist a location $k$ that never sends signal $0$ and which the customer always finds optimal to join, regardless of other locations' signals. This case corresponds to condition (I).  Alternatively, all locations send both signals with positive probability; in this case, optimality of $f \in \mathcal{F}_d$ suggests that the customer finds it optimal to join a location if it is the only one sending signal $1$, and finds it optimal to not join any location if all locations send signal $0$. These requirements are captured in condition (II). The proof in Section~\ref{sec:proofs}  shows that these necessary conditions are in fact sufficient.

The preceding two lemmas allow us to completely eliminate the customer's strategy from the problem of identifying the optimal decentralized signaling mechanism, and pose it solely in terms of the signaling mechanism. However, the system's objective function in the formulation~\eqref{eq:sys_org_obj} is still a non-linear expression, as shown in Lemma~\ref{lem:thput-ind-fd}. Nevertheless, as we show in the next section, the simplification achieved from these results allows us to provide an efficient way to solve the problem.

\subsection{Isolated system LP formulation}

In this section, we show that to solve the non-linear decentralized signaling problem for $\sys$, we may simply solve $K$ linear programs, corresponding to $K$ single-location systems $\mathsf{SYS}_1(\mu_k,h_k)$ for $k \in \comp$. For each $k \in \comp$, we introduce the notion of $\iso$, which is the single-location system $\mathsf{SYS}_1(\mu_k,h_k)$ consisting of a single location $k$ with state distribution $\mu_k$. Applying the formulation~\eqref{eq:centralized-lp} to the $\iso$, we obtain that the optimal throughput $\thput^{\mathsf{iso}}_{k}$ of $\iso$ can be found by solving the following linear program:
\begin{align}
\thput^{\mathsf{iso}}_{k}= \max_{\sigma\in\mathcal{C}(\{0, 1\})}  &\sum_{\omega_k \in \Omega_k} \mu_k(\omega_k) \sigma(1|\omega_k)  \notag\\
\text{s.t. } &\sum_{\omega_k \in \Omega_k}\mu_k(\omega_k) \sigma(1|\omega_k)h_k(\omega_k) \geq 0\notag\\
&\sum_{\omega_k \in \Omega_k}\mu_k(\omega_k) \sigma(0|\omega_k)h_k(\omega_k) \leq 0.\label{eq:iso-lp}
\end{align}
In Theorem~\ref{thm:decentralized_isolated_equivalence}, we show that an optimal decentralized signaling mechanism can be constructed for $\sys$ using optimal solutions of the $K$ linear programs~\eqref{eq:iso-lp}.

\begin{theorem}\label{thm:decentralized_isolated_equivalence}
For each $k \in \comp$, let $\sigma_k \in \mathcal{C}(\{0,1\})$ be an optimal obedient signaling mechanism for $\iso$. Then, the signaling mechanism $\sigma \in \mathcal{D}(\uset)$ with $\sigma(s|\omega)= \prod_{k \in \comp} \sigma_k(s_k|\omega_k)$ for all $s \in \uset$ and $\omega \in \Omega$ is an optimal decentralized signaling mechanism for $\sys$. In particular, there exists a strategy $f \in \mathcal{F}_d$ that is optimal under $\sigma$ (i.e., satisfying \eqref{eq:optimal-strategy}) with
\begin{align*}
    \thputD = T(\sigma, f) &= 1 - \prod_{k \in \comp} (1 - \thput^{\mathsf{iso}}_k).
\end{align*}
\end{theorem}
\proof{Proof.} 
For each $k \in \comp$, let $\sigma_k \in \mathcal{C}(\{0, 1\})$ be an optimal obedient signaling mechanism for $\iso$. Define the signaling mechanism $\sigma \in \mathcal{D}(\uset)$ with $\sigma(s|\omega) = \prod_{k\in \comp} \sigma_k(s_k | \omega_k)$ for $s \in \uset$ and $\omega \in \Omega$. Since each $\sigma_k$ is feasible for $\iso$, we conclude by condition (II) of Lemma~\ref{lem:decentralized-obedience} that there exists an $f \in \mathcal{F}_d$ that is optimal under $\sigma$. Furthermore, from Lemma~\ref{lem:thput-ind-fd}, we obtain that 
\begin{align*}
    T(\sigma, f) &= 1 - \prod_{k \in \comp}\left( 1 - \sum_{\omega_k \in \Omega_k} \mu_k(\omega_k) \sigma_k(1 | \omega_k)\right)\\
    &= 1 - \prod_{k \in \comp}\left( 1 - \thput_k^{\mathsf{iso}}\right).
\end{align*}

Let $\sigma' \in \mathcal{D}(\uset)$ be any other signaling mechanism for which there exists an $f' \in \mathcal{F}_d$ that is optimal. Then, $\sigma'$ satisfies one of the two conditions in Lemma~\ref{lem:decentralized-obedience}. 

Suppose $\sigma'$ satisfies condition (I) in Lemma~\ref{lem:decentralized-obedience}. Then, there exists an $\ell$ such that $\mu_\ell(\omega_\ell) \sigma'_\ell(0|\omega_\ell) = 0$ for all $\omega_\ell \in \Omega_\ell$, and $\sum_{\omega_\ell \in \Omega_\ell} \mu_\ell(\omega_\ell) h_\ell(\omega_\ell) \geq 0$. From these, we conclude that $\thput_\ell^{\mathsf{iso}} = 1$. This, in turn implies that $T(\sigma, f) = 1 \geq T(\sigma', f')$. 

Suppose instead that $\sigma'$ satisfies condition (II) in the Lemma~\ref{lem:decentralized-obedience}. Then, for all $k \in \comp$, $\sigma_k'$ is feasible for  $\mathsf{ISO}(\mu_k, h_k)$. By the optimality of $\sigma_k$ for  $\mathsf{ISO}(\mu_k, h_k)$, we obtain,  
\begin{align*}
    \sum_{\omega_k \in \Omega_k} \mu_k(\omega_k) \sigma'_k(1|\omega_k) \leq \sum_{\omega_k \in \Omega_k} \mu_k(\omega_k)\sigma_k(1|\omega_k).
\end{align*} 
Thus, using Lemma~\ref{lem:thput-ind-fd}  and the fact that $f' \in \mathcal{F}_d$, we get
\begin{align*}
T(\sigma', f') &= 1- \prod_{k \in \comp} \left( 1 - \sum_{\omega_k \in \Omega_k} \mu_k(\omega_k) \sigma'_k(1| \omega_k)\right)\\ 
&\leq 1- \prod_{k \in \comp} \left( 1 - \sum_{\omega_k \in \Omega_k} \mu_k(\omega_k) \sigma_k(1| \omega_k)\right)\\ 
&= T(\sigma, f).
\end{align*}

Thus, under either condition, we conclude that $\sigma$ is an optimal decentralized signaling mechanism for $\sys$.
\Halmos\endproof

Having analyzed both centralized and decentralized signaling problems, we are ready to return in the next section to the problem of understanding how well decentralized signaling mechanisms perform compared to centralized signaling mechanisms.  

\section{A guarantee for decentralized signaling}
\label{sec:cost-dec}

In this section we establish that for any $K$-location system as described in Section~\ref{sec:system}, the best decentralized signaling mechanism yields a throughput that is at least a strictly positive constant multiple of the throughput of the best centralized signaling mechanism.  The constant depends only upon $K$ and not on other system properties such as state distributions or customer utility functions.  The following theorem is the main result of this section.
\begin{theorem}\label{thm:cost-of-decentralization} For the system $\sys$, we have $\thputD \geq \left[1 - \left(1-\frac{1}{K}\right)^K\right]  \thput$. 
\end{theorem}
Before we present the proof, we provide a few comments about the theorem. First, as noted above, the result holds uniformly, that is, for any $K$-location system with independent states. For instance, the theorem implies that $\thputD \geq \frac{3}{4} \thput$ for {\em any} such system with two locations and that $\thputD \geq \frac{19}{27} \thput$ for {\em any} such system with three locations.  The term $1- \left(1-1/K\right)^K$ decreases monotonically to $1-e^{-1}$ as $K$ increases. Therefore, we can obtain a performance guarantee that does not depend upon the number of locations $K$; in particular, it holds that $\thputD \geq [1-e^{-1}] \thput \ge 0.63 \cdot \thput$ for any system with any number of independent locations. 
\proof{Proof.} Let $\sigma \in \mathcal{C}(\comp_0)$ be an optimal centralized signaling mechanism, i.e., an optimal solution to \eqref{eq:centralized-lp}. Let $T_k(\sigma)$ denote the throughput under $\sigma$ (and under the obedient strategy) through location $k \in \comp$, defined as
\begin{align*}
    T_k(\sigma) \coloneq \sum_{\omega \in \Omega} \mu(\omega) \sigma(k|\omega). 
\end{align*}
Given the optimality of $\sigma$ to the problem~\eqref{eq:centralized-lp}, we have $\thput = \sum_{k \in \comp} T_k(\sigma)$. 

We will now argue that $\thput_k^{\mathsf{iso}} \ge T_k(\sigma)$ for $k \in \comp$.
For this, fix $k \in \comp$, and define a signaling mechanism $\sigma_k \in \mathcal{C}(\{0,1\})$ for the $\iso$ system as follows: for all $\omega_k \in \Omega_k$,
\begin{align*}
    \sigma_k(1|\omega_k) &\coloneq \sum_{\omega_{-k} \in \Omega_{-k}} \mu_{-k}(\omega_{-k}) \sigma(k | \omega_k, \omega_{-k}),    
\end{align*}
and $\sigma_k(0|\omega_k) = 1 - \sigma_k(1|\omega_k)$. Suppose, in the $\iso$ system, the signaling mechanism $\sigma_k$ is adopted. If $\prob^{\sigma_k}(\hat{u}_k  = 1) = 0$, then $T_k(\sigma) = 0 \leq \thput_k^{\mathsf{iso}}$. If $\prob^{\sigma_k}(\hat{u}_k  = 1) > 0$, then the expected utility of a customer for joining the location after receiving signal $\hat{u}_k = 1$ is given by
\begin{align*}
    &\expec^{\sigma_k}\left[ h_k(\hat{\omega}_k) | \hat{u}_k = 1\right] \\
    &\quad = \frac{1}{\prob^{\sigma_k}(\hat{u}_k  = 1)} \expec^{\sigma_k}\left[ h_k(\hat{\omega}_k) \ind\{ \hat{u}_k = 1\}\right] \\
    &\quad = \frac{1}{\prob^{\sigma_k}(\hat{u}_k  = 1)} \sum_{\omega_k \in \Omega_k} \mu_k(\omega_k) \sigma_k(1|\omega_k) h_k(\omega_k)\\
    &\quad = \frac{1}{\prob^{\sigma_k}(\hat{u}_k  = 1)} \sum_{\omega_k \in \Omega_k} \sum_{\omega_{-k} \in \Omega_{-k}} \mu_k(\omega_k) \mu_{-k}(\omega_{-k}) \sigma(k | \omega_k, \omega_{-k}) h_k(\omega_k)\\
    &\quad = \frac{1}{\prob^{\sigma_k}(\hat{u}_k  = 1)} \sum_{\omega \in \Omega} \mu(\omega) \sigma(k | \omega) h_k(\omega_k)\\
    &\quad \geq 0.
\end{align*}
Here, the third equality follows from the definition of $\sigma_k$, and final inequality follows from the fact that $\sigma$ is optimal (and hence feasible) for \eqref{eq:centralized-lp}. Thus, we conclude that there exists a strategy $f_k \in \mathcal{F}(\{0, 1\})$ satisfying $f_k(1|1) = 1$ that is optimal under $\sigma_k$ for the $\iso$ system. Let $T_k(\sigma_k, f_k)$ be the throughput in the $\iso$ system under signaling mechanism $\sigma_k$ and customer strategy $f_k$. Hence, we obtain 
\begin{align*}
    \thput_k^{\mathsf{iso}} &\geq T_k(\sigma_k, f_k) \\
    &= \sum_{\omega_k \in \Omega_k} \mu_k(\omega_k) \sum_{u_k \in \{0, 1\}} \sigma_k(u_k | \omega_k) f_k(1|u_k) \\
    &\geq \sum_{\omega_k \in \Omega_k} \mu_k(\omega_k) \sigma_k(1 | \omega_k) \\
    &= \sum_{\omega_k \in \Omega_k} \mu_k(\omega_k) \sum_{\omega_{-k} \in \Omega_{-k}} \mu_{-k}(\omega_{-k}) \sigma(k | \omega_k , \omega_{-k})\\
    &= \sum_{\omega \in \Omega} \mu(\omega) \sigma(k | \omega)\\
    &= T_k(\sigma).
\end{align*}
Here, the first inequality follows from the definition of $\thput_k^{\mathsf{iso}}$, the second inequality follows because $f_k(1|1)=1$, $f_k \geq 0$ and $\sigma_k \geq 0$, and the final equality follows from the definition of $T_k(\sigma)$. 

To complete the proof of the theorem, we have from Theorem~\ref{thm:decentralized_isolated_equivalence} that
\begin{align*}
    \thputD &= 1 - \prod_{k \in \comp} \left(1 - \thput_k^{\mathsf{iso}}\right) \\
     &\geq 1 - \prod_{k \in \comp} \left(1 - T_k(\sigma) \right) \\
    &\geq \left[ 1 - \left(1- \frac{1}{K}\right)^K\right] \cdot \sum_{k \in \comp} T_k(\sigma) \\
    &= \left[ 1 - \left(1- \frac{1}{K}\right)^K\right] \cdot \thput.
\end{align*}
Above, the first inequality follows because $\thput_k^{\mathsf{iso}} \geq T_k(\sigma)$ for each $k$. The second inequality follows from Lemma~\ref{lem:series_algebra} in Section~\ref{sec:proofs}. 
The final equality follows from the optimality of $\sigma$ for the centralized problem.
\Halmos\endproof

We close this section with the following theorem, which shows that the bound obtained in Theorem~\ref{thm:cost-of-decentralization} is tight. In other words, one cannot make the constant $1 - \left(1-1/K\right)^K$ in Theorem~\ref{thm:cost-of-decentralization} larger and still have the inequality there hold for all $K$-location systems.

\begin{theorem}\label{thm:lower-bound} For any $\epsilon > 0$ and $K \geq 2$, there exists a $K$-location system $\sys[(\mu, h_1, \dots, h_K)]$ for which $\thputD \leq \left[ 1 - \left(1 - \frac{1+\epsilon}{K}\right)^K\right] \thput$.
\end{theorem}
\proof{Proof.} Consider a system $\sys$ with $K$ locations, where the state of each location $k$ takes values in $\Omega_k = \{0, 1\}$ with $\mu_k(1)=p$. For each $k \in \comp$, let $h_k(\omega_k) = X \ind\{ \omega_k = 1\} + (-1) \ind\{\omega_k = 0\}$ for some large enough $X > 1$. This is the same system we considered in Example~\ref{ex:inadequacy}. Recall from the analysis there that if $p = p^* \coloneq 1 - \left(\frac{X}{X+1}\right)^{1/K}$, then $\thput = 1$.

Next, we use Theorem~\ref{thm:decentralized_isolated_equivalence} to identify the optimal decentralized signaling mechanism. Consider the isolated system $\iso$ for $k \in \comp$. By analyzing the LP~\eqref{eq:iso-lp}, it is straightforward to show that the signaling mechanism $\sigma_k \in \mathcal{C}(\{0,1\})$ defined below is optimal:
\begin{align*}
    \sigma_k(1 | \omega_k) &\coloneq \begin{cases} 1 & \text{if $\omega_k = 1$;}\\
    z^* & \text{if $\omega_k = 0$,}
    \end{cases}
\end{align*}
where $z^* \coloneq \frac{p^* X}{1 - p^*}$ is chosen as the maximal value for which the obedience conditions hold; in particular, upon receiving the signal $1$, a customer obtains expected utility for joining equal to $\frac{p^* X+ (1-p^*)z^*(-1)}{p^* + (1- p^*)z^*} = 0$. Under $\sigma_k$, the probability of joining location $k$ is given by $p^* + (1 - p^*)z^*$. Since this is true for all $k \in \comp$, by Theorem~\ref{thm:decentralized_isolated_equivalence}, we obtain
\begin{align*}
    \thputD &= 1 - \left(1 - [p^* + (1-p^*)z^*]\right)^K\\
    &= 1 - \left( 1 - p^* (X+1)\right)^K.
\end{align*}
Since $\thput = 1$, the theorem statement will follow upon showing that, for any $\epsilon > 0$, there exists an $X$ large enough that $p^*(X+1) < \frac{1+\epsilon}{K}$. To see that such $X$ does indeed exist, observe that
\begin{align*}
    \lim_{X \to \infty} p^* (X+1) =  
    \lim_{X \to \infty} \frac{ 1 - \left(\frac{X}{X+1}\right)^{1/K}}{\frac{1}{X+1}} = \frac{1}{K},
\end{align*}
where the last equality follows from L'H\^{o}pital's Rule.
\Halmos\endproof

\section{Heterogeneous preferences over locations}\label{sec:heterogeneous}

We have heretofore assumed that the system operator is indifferent to which location a customer joins.  In this section, we relax this assumption and instead 
suppose that the system operator has heterogeneous preferences over the locations. In particular, we assume that if the customer joins a location $k \in \comp$, the system operator receives a payoff equal to $v_k$; we continue to assume that the payoff for not joining any location equals zero. The setting we analyzed until now corresponds to the case where $v_k = 1$ for all $k \in \comp$. Note that we do not assume that the $v_k$'s are positive. 

Analogous to \eqref{eq:thput}, we define the system operator's value from using a signaling mechanism $\sigma \in \mathcal{C}(\sign)$, if the customer adopts a strategy $f \in \mathcal{F}(\sign)$, as follows:
\begin{align*}
    V(\sigma, f) &\coloneq \expec^{\sigma, f} \left[ \sum_{a \in \comp}  \ind\{\hat{a} = a\} v_a \right] = \sum_{\omega \in \Omega} \sum_{s \in \sign} \sum_{a \in \comp} \mu(\omega) \sigma(s|\omega) f(a|s) v_a.
\end{align*}

Denote the value of the optimal centralized signaling mechanism as $\val$:
\begin{align*}
    \val \coloneq \max_{\sign} \max_{\sigma \in \mathcal{C}(\sign), f \in \mathcal{F}(\sign)} V(\sigma, f) \quad \text{s.t. $f$ satisfies \eqref{eq:optimal-strategy} under $\sigma$.}
\end{align*} 
Analogously, denote by $\valD$ the value of the optimal decentralized signaling mechanism, obtained by restricting the above optimization to $\sigma \in \mathcal{D}(\sign)$ with $\sign = \sign_1 \times \dots \times \sign_K$.

Just as for the case of throughput, the revelation principle can be used to show that the optimal centralized signaling mechanism can be found by solving a linear program analogous to~\eqref{eq:centralized-lp}. For the decentralized setting, we can reduce the system operator's problem to a non-linear formulation analogous to~\eqref{eq:decentralized-nlp} through a similar argument. However, our subsequent analysis of the decentralized setting and its reduction to the $K$-linear programs in~\eqref{eq:iso-lp} heavily used the fact that the system operator is indifferent among the locations. This argument fails in the case of heterogeneous preferences, where, in addition to the question of whether the customer joins a location, it is important to address which location the customer joins. 

Despite this lack of tractability, as we show next, our analysis of the isolated systems $\iso$ continues to provide similar bounds on the performance of the optimal decentralized mechanism under the following assumption: 
\begin{assumption}\label{as:negative-util}
    For each location $k \in \comp$, we have $\expec[ h_k(\hat{\omega}_k)] < 0$. 
\end{assumption}
The preceding assumption implies that, if a customer receives no information about the state of a location $k \in \comp$, then the customer will not find it optimal to join that location. Thus, some level of information sharing is needed to convince the customer to join some location.
The following result states that, under the assumption, the previous section's lower bound on the performance of the optimal decentralized mechanism applies in the heterogeneous preferences setting as well.

\begin{theorem}\label{thm:cost-dec-heterogeneous} Suppose Assumption~\ref{as:negative-util} holds. Then, there exists a decentralized signaling mechanism $\sigma \in \mathcal{D}(U)$ and a strategy $f \in \mathcal{F}_d$, such that $f$ is optimal under $\sigma$, and 
\begin{align*}
     V(\sigma, f) \geq \left[ 1 - \left(1 - \frac{1}{K}\right)^K\right]  \val. 
\end{align*}
In particular, we obtain $\valD \geq \left[1 - \left(1 - K^{-1}\right)^K\right] \val$. 
\end{theorem}
\proof{Proof.} First, observe that a customer can never be persuaded  (by a centralized or a decentralized signaling mechanism) to join a location $k$ for which $h_k(\omega_k) < 0$ for all $\omega_k \in \Omega_k$ with $\mu_k(\omega_k) > 0$. Inclusion of such locations into the set of locations only increases $K$ and makes the desired bound weaker. Thus, without loss of generality, assume that for each location $k$, there exists a state $\bar{\omega}_k \in \Omega_k$ with $\mu_k(\bar{\omega}_k) > 0$ and $h_k(\bar{\omega}_k) \geq 0$. 

Similarly, under Assumption~\ref{as:negative-util}, any location $k$ with $v_k \leq 0$ can be safely ignored: a centralized mechanism will never ask a customer to join that location, and a decentralized mechanism can also achieve this outcome by ensuring that the location never reveals any information. As before, inclusion of such locations only makes the desired bound weaker. Thus, without loss of generality, we assume that $v_1 \geq v_2 \geq \dots \geq v_K > 0$. For convenience, we define $v_{K+1} = 0$.

Let $\psi \in \mathcal{C}(\comp_0)$ be an optimal centralized signaling mechanism (with the obedient strategy being optimal under $\psi$). As in the proof of Theorem~\ref{thm:cost-of-decentralization}, let $T_k(\psi)$ denote the throughput under $\psi$ (and under the obedient strategy) through location $k \in \comp$: $T_k(\psi) = \sum_{\omega \in \Omega} \mu(\omega) \psi(k|\omega)$. Note that the optimality of $\psi$ implies that $\val = \sum_{k \in \comp} v_k T_k(\psi)$. 

Next, for each $k \in \comp$, let $\sigma_k \in \mathcal{C}(\{0, 1\})$ be the optimal signaling mechanism for the $\iso$ system, i.e., $\sigma_k$ is the solution to the LP~\eqref{eq:iso-lp}. Under Assumption~\ref{as:negative-util} and our earlier assumption that there exists an $\bar{\omega}_k \in \Omega_k$ with $\mu_k(\bar{\omega}_k) > 0$ and $h_k(\bar{\omega}_k) \geq 0$, one can show by analyzing the LP~\eqref{eq:iso-lp} that (i) $\sum_{\omega_k \in \Omega_k} \mu_k(\omega_k) \sigma_k(1|\omega_k) h_k(\omega_k) = 0$ and $\sum_{\omega_k \in \Omega_k} \mu_k(\omega_k) \sigma_k(0|\omega_k) h_k(\omega_k) < 0$ and (ii) $\thput_k^\mathsf{iso} = \prob^{\sigma_k}(\hat{u}_k = 1) = \sum_{\omega_k \in \Omega_k} \mu_k(\omega_k) \sigma_k(1|\omega_k) \in (0,1)$.

Let $\sigma \in \mathcal{D}(\uset)$ be the mechanism with $\sigma(u|\omega) = \prod_{k \in \comp} \sigma_k(u_k|\omega_k)$ for all $u \in \uset$ and $\omega \in \Omega$. By the independence of the locations and the discussion in the preceding paragraph, it follows that for all $u \in \uset$, we have $\prob^{\sigma}(\hat{u} = u) > 0$. Also by the discussion in the previous paragraph, for each $u \in \uset$,  we have $\expec^\sigma[ h_k(\hat{\omega}_k) | \hat{u} = u] = 0$ for any location $k$ with $u_k = 1$, and $\expec^\sigma[ h_k(\hat{\omega}_k) | \hat{u} = u] < 0$ for any location $k$ with $u_k = 0$. Thus, upon receiving a signal vector $u \in \uset$ from $\sigma$, the customer is indifferent among joining any location with corresponding signal $1$, and strictly prefers not to join a location with corresponding signal $0$.

Let $f \in \mathcal{F}(\uset)$ be defined as follows: for $u \in \uset$ and $k \in \comp$, 
\begin{align}\label{eq:tie-breaker}
    f(k|u) \coloneq \begin{cases} 
    1 & \text{if $u_k = 1$ and $u_\ell = 0$ for all $\ell < k$;}\\
    0 & \text{otherwise,}
    \end{cases}
\end{align}
with $f(0|u) = 1 - \sum_{k\in \comp} f(k|u)$. In particular, among all locations that send signal $1$, the strategy $f$ picks the location with the smallest index. It is straightforward to verify that $f \in \mathcal{F}_d$. By the discussion in the preceding paragraph, it further follows that $f$ is optimal under $\sigma$. 

Let $\Gamma_k \coloneq 1 - (1 - 1/k)^k$ for $k \in \comp$. Note that $\Gamma_k$ is decreasing in $k$. Also, let $\zero_k$ denote a vector of $k$ zeros, and $\hat{u}_{1:k}$ denote the first $k$ elements of the signal vector $\hat{u}$. We now have
\begin{align*}
    V(\sigma, f) &= \expec^{\sigma, f}\left[ \sum_{k \in \comp} \ind\{\hat{a} = k\} v_k \right]\\
    &= \expec^\sigma \left[ \ind\{\hat{u}_1 = 1\} v_1 + \sum_{k =2}^K \ind\{ \hat{u}_{1:{k-1}} = \zero_{k-1}, \hat{u}_k = 1 \} v_k\right]\\
    &= \expec^\sigma \left[ (1 - \ind\{\hat{u}_1 = 0\}) v_1 + \sum_{k=2}^K \left( \ind\{ \hat{u}_{1:{k-1}} = \zero_{k-1} \}  - \ind\{ \hat{u}_{1:k}  = \zero_k \} \right) v_k\right]\\
    &= \expec^\sigma\left[ \sum_{k =1}^K (v_k - v_{k+1})\left( 1 -  \ind\{ \hat{u}_{1:k}  = \zero_k \}\right) \right]\\ 
    &= \sum_{k =1}^K (v_k - v_{k+1}) \left(1 - \prob^\sigma\left( \hat{u}_{1:k}  = \zero_k \right)\right)\\ 
    &= \sum_{k =1}^K (v_k - v_{k+1}) \left[1 - \prod_{\ell =1}^k \left(1 - \thput_\ell^{\mathsf{iso}}\right)\right]\\ 
    &\geq \sum_{k =1}^K (v_k - v_{k+1}) \left[ 1 -  \prod_{\ell =1}^k \left(1 - T_\ell(\psi)\right) \right]\\ 
    &\geq \sum_{k=1}^K  (v_k - v_{k+1}) \left[ \Gamma_k \sum_{\ell =1}^k T_\ell(\psi)\right]\\
    &\geq \Gamma_K \sum_{k=1}^K  (v_k - v_{k+1})  \sum_{\ell =1}^k T_\ell(\psi)\\
    &= \Gamma_K \cdot \sum_{k=1}^K v_k T_k(\psi)\\
    &= \Gamma_K \cdot \val.
\end{align*}
Here, we have used the definition~\eqref{eq:tie-breaker} of $f$ in the second equality; the sixth equality uses the independence of the signals under the decentralized mechanism $\sigma$ and the fact that $\prob^{\sigma_k}(\hat{u}_k = 0) = 1 - \thput_k^{\mathsf{iso}}$ due to optimality of $\sigma_k$ to \eqref{eq:iso-lp}. The first inequality follows from the fact that,  for all $k \in \comp$,  $\thput_k^{\mathsf{iso}} \geq T_k(\psi)$, as shown in the proof of Theorem~\ref{thm:cost-of-decentralization}, and because $v_k \geq v_{k+1}$. In the second inequality, we have used Lemma~\ref{lem:series_algebra}, and in the third inequality, we use $\Gamma_k \geq \Gamma_K$. Since $\valD \geq V(\sigma, f)$, we obtain the theorem statement.   
\Halmos\endproof

Note that the example in Theorem~\ref{thm:lower-bound} also shows that this bound is tight, since Assumption~\ref{as:negative-util} holds in that example. We leave the question of whether our lower bound is achievable without Assumption~\ref{as:negative-util} for future work. 

\section{Correlated locations}
\label{sec:correlated}
Until now, we have assumed that the states of different locations are independent. In this section, we relax this assumption of independence and allow the states to have an arbitrary joint distribution. Without independence, the information that one location reveals affects not only the customer's belief about that location's state, but also the belief about the states of all other locations. Consequently, our approach of analyzing each location in isolation with the $\iso$ system is no longer workable, and a completely different approach is needed. 
Our first main result of this section is the following bound on the throughput of an optimal decentralized signaling mechanism.

\begin{theorem} 
\label{thm:correlated}
Suppose the locations' states are distributed according to an arbitrary joint distribution $\mu$. Then, we have
\begin{align*}
    \thputD \geq \frac{1}{K} \thput.
\end{align*}    
\end{theorem}

Before proving the theorem, it is instructive to make comparisons with Theorem~\ref{thm:cost-of-decentralization}, which applies in settings where the locations' states are independent. As noted above and in the introduction, the guarantee in Theorem~\ref{thm:correlated} is weaker than the guarantee in Theorem~\ref{thm:cost-of-decentralization}.  For a system with two locations ($K=2$), Theorem~\ref{thm:cost-of-decentralization} ensures that decentralized signaling yields at least 75\% of the throughput of centralized signaling in settings with independent locations. Without the assumption of independence, Theorem~\ref{thm:correlated} provides a guarantee of just 50\%.
The difference in guarantees is quite significant when $K$ is large, that is, when there are many locations.  As $K \to \infty$, Theorem~\ref{thm:cost-of-decentralization} guarantees that decentralized signaling yields at least 63\% of the throughput of centralized signaling.  On the other hand, as $K \to \infty$ the bound in Theorem~\ref{thm:correlated} approaches zero. This suggests that in the presence of dependence across the locations' states, there is considerable benefit to the coordination built into centralized signaling.  One might wonder whether this is simply an artifact of the proof of Theorem~\ref{thm:correlated} and whether better guarantees are possible for correlated locations. As we will see in Theorem~\ref{thm:correlated-upper-bound} below, there can at best be minimal improvements to the dependence on $K$ found in Theorem~\ref{thm:correlated}.

\proof{Proof.} Let $\sigma \in \mathcal{C}(\comp_0)$ be an optimal centralized signaling mechanism, i.e., $\sigma$ is the solution to \eqref{eq:centralized-lp} (with $\mu$ now an arbitrary distribution). As in the proof of Theorem~\ref{thm:cost-of-decentralization}, let $T_k(\sigma)$ denote the throughput under $\sigma$ (and under the obedient strategy) through location $k \in \comp$, and note that the optimality of $\sigma$ implies  $\thput = \sum_{k \in \comp}  T_k(\sigma)$. 

To show our lower bound, we will construct a decentralized signaling mechanism $\psi$, where all locations, except for a carefully selected location $k$, reveal no information, and location $k$ shares a binary signal $\hat{u}_k$, representing the recommendation to join (or not join). This binary signal is obtained by location $k$ simulating the states of all other locations (given its own state), generating a signal $\hat{s}$ according to $\sigma$ given that (partially simulated) state vector, and recommending the customer to join itself if and only if $\hat{s} = k$. We now describe this construction more formally.

Let $k \in \comp$ be a location for which we have $T_k(\sigma) \geq \frac{1}{K} \sum_{\ell \in \comp}  T_\ell(\sigma) = \frac{1}{K} \thput$.  Define $\psi_k \in \mathcal{C}(\{0,1\})$ as follows: for all $\omega_k \in \Omega_k$, let
\begin{align*}
     \psi_k(1 | \omega_k) &\coloneq \sum_{\omega_{-k} \in \Omega_{-k}} \mu_{-k}(\omega_{-k} | \omega_k) \sigma(k | \omega_{-k}, \omega_k),
\end{align*}
and $\psi_k(0|\omega_k) = 1 - \psi_k(1|\omega_k)$. Here, $\mu_{-k}(\cdot | \omega_k)$ denotes the conditional distribution of the states of all locations other than $k$, conditional on $\hat{\omega}_k = \omega_k$. (If $\prob(\hat{\omega}_k = \omega_k) = 0$, then define $\mu_{-k}$ arbitrarily.) For $\ell \neq  k$, define $\psi_\ell \in \mathcal{C}(\{0,1\})$ as $\psi_\ell(u_\ell |  \omega_\ell) \coloneq \ind\{u_\ell = 0\}$ for $u_\ell \in \{0,  1\}$ and for all $\omega_\ell \in \Omega_\ell$. Note that $\psi_\ell$ reveals no information about $\hat{\omega}$ because it always sends a signal of $0$. Finally, define the decentralized signaling mechanism $\psi \in \mathcal{D}(\uset)$ as $\psi(u | \omega) \coloneq \prod_{\ell \in \comp} \psi_\ell(u_\ell | \omega_\ell)$ for $u \in \uset$ and $\omega \in \Omega$. 

Note that under $\psi$, only two possible signal vectors can arise with positive probability: the signal vector $\zero$ (where all locations send signal $0$) and the signal vector $(\zero_{-k}, 1)$ (where all locations other than $k$ send signal $0$ and location $k$ sends signal $1$). 

If $\prob^{\psi}(\hat{u} = (\zero_{-k}, 1)) = 0$, then it follows that $T_k(\sigma) = 0$, and hence $\thput = 0$, and theorem statement holds trivially. Thus, assume $\prob^{\psi}(\hat{u} = (\zero_{-k}, 1)) > 0$. We have
\begin{align*}
    &\expec^{\psi} \left[ h_k(\hat{\omega}_k) | \hat{u} = (\zero_{-k}, 1)\right]\\
    &\quad = \expec^{\psi_k} \left[ h_k(\hat{\omega}_k) | \hat{u}_k = 1\right]\\
    &\quad = \frac{1}{\prob^{\psi_k}\left( \hat{u}_k = 1\right)} \expec^{\psi_k} \left[ h_k(\hat{\omega}_k) \ind\{\hat{u}_k = 1\} \right]\\
    &\quad = \frac{1}{\prob^{\psi_k}\left( \hat{u}_k = 1\right)} \sum_{\omega_k \in \Omega_k} \mu_k(\omega_k) \psi_k(1 | \omega_k) h_k(\omega_k)\\
    &\quad = \frac{1}{\prob^{\psi_k}\left( \hat{u}_k = 1\right)} \sum_{\omega_k \in \Omega_k} \mu_k(\omega_k) \sum_{\omega_{-k} \in \Omega_{-k}} \mu_{-k}(\omega_{-k} | \omega_k) \sigma(k | \omega_{-k}, \omega_k) h_k(\omega_k)\\
    &\quad = \frac{1}{\prob^{\psi_k}\left( \hat{u}_k = 1\right)} \sum_{\omega \in \Omega} \mu(\omega) \sigma(k|\omega) h_k(\omega_k)\\
    &\quad \geq 0.
\end{align*}
where the inequality follows from the fact that the obedient strategy is optimal under $\sigma$. Thus, we conclude that, under $\psi$, there exists an optimal strategy $f$ where the customer joins a location upon receiving the signal $\hat{u} = (\zero_{-k}, 1)$. (Note that it is not necessary that the customer joins location $k$; due to dependencies across locations, the customer may find it preferable to join some other location based on the information revealed by location $k$.) Thus, we obtain
\begin{align*}
    T(\psi, f ) &\geq \prob^{\psi}\left( \hat{u}  = (\zero_{-k}, 1)\right) \\
    &= \sum_{\omega_k \in \Omega_k} \mu_k(\omega_k) \psi(1|\omega_k)\\
    &= \sum_{\omega_k \in \Omega_k} \mu_k(\omega_k) \sum_{\omega_{-k} \in \Omega_{-k}} \mu_{-k}(\omega_{-k} | \omega_k) \sigma(k | \omega_{-k}, \omega_k) \\
    &= \sum_{\omega \in \Omega} \mu(\omega) \sigma(k|\omega)\\
    &= T_k(\sigma) \\
    &\geq \frac{1}{K} \thput.
\end{align*}
This completes the proof.
\Halmos\endproof

The preceding bound, while positive for any fixed $K$, vanishes as $K \to \infty$. This is in contrast to the independent locations setting of Theorem~\ref{thm:cost-of-decentralization}, where there is a positive lower bound on throughput independent of $K$. Is it possible to obtain a positive multiplicative factor guarantee for correlated locations, just as in the case of independent locations? The analysis below shows that in general, the answer to this question is no; in fact, the bound in Theorem~\ref{thm:correlated} cannot be substantially improved.

For this, consider a $K$-location system with $\Omega_k = \{0, \pm 1\}$ for each $k \in \comp$. Let $h_k(\omega_k) = K \ind\{ \omega_k = 1\} + (-1) \ind\{ \omega_k = 0\} + (-X) \ind\{\omega_k = -1\}$, where $X \gg K$. Define the distribution $\mu$ as follows:
    \begin{align*}
      \mu(\omega) &= \begin{cases} \frac{1}{K} \cdot \frac{1}{K+1} & \text{if $\omega$ is a permutation of $(1, \zero_{k-1})$;}\\
        \frac{1}{K} \cdot \frac{K}{K+1}  & \text{if $\omega$ is a permutation of $(0, -\one_{k-1})$;}\\
        0 & \text{otherwise.}
      \end{cases}
    \end{align*}
    In particular, under $\mu$, with probability $1/(K+1)$, a
    location, chosen uniformly at random, is in state $1$ and all
    other locations are in state $0$, whereas with remaining
    probability $K/(K+1)$, a location, again chosen uniformly at
    random, is in state $0$ and all other locations are in state
    $-1$. It is straightforward to verify that, for all $X > K$, we
    have $\thput = 1$, which is achieved by a centralized signaling
    mechanism that always asks the customer to join the unique
    location in state $1$ if any such location exists, and if not,
    asks the customer to join the unique location in state $0$.

We are now ready to prove the following theorem, which bounds the
    throughput of any decentralized signaling mechanism in the above system under the assumption that the customer receives utility of $-\infty$ if she joins a location that is in state~$-1$.
    An immediate consequence of the theorem and the above discussion is that $\thputD / \thput = \mathcal{O}(\log K / K)$ in this example. Thus, as mentioned earlier, the bound in Theorem~\ref{thm:correlated} cannot be improved much.
    
    \begin{theorem}\label{thm:correlated-upper-bound} Consider the $K$-location
      system described above with $X = \infty$. Let
      $\sign = \sign_1\times \dots \times \sign_K$ for finite sets $\{\sign_k : k \in \comp\}$, and let $\sigma \in \mathcal{D}(\sign)$
      be a decentralized signaling mechanism. Let $z^*_K$ be the unique solution in $[0,1]$ to $z - (1-z)^{K-1}=0$. Then,
      for any strategy $f$ that is optimal under $\sigma$, we have
      $T(\sigma, f) \leq \frac{1 + Kz^*_K}{1+K} = \mathcal{O}(\log K / K)$.
    \end{theorem}

\proof{Proof.}
Define $\hat{\theta} \in \{0, 1\}$
  as follows: $\hat{\theta} = 1$ if and only if
  $\hat{\omega}$ is a permutation of $(1, \zero_{K-1})$. Note that, under $\mu$, we
  have $\prob(\hat{\theta} = 1) = \frac{1}{K+1}$. Let $\mu_\theta$
  denote the conditional distribution of $\hat{\omega}$, given
  $\hat{\theta} = \theta \in \{0, 1\}$. We observe that $\mu_1$ is a uniform distribution over all permutations of $(1, \zero_{K-1})$ and $\mu_0$ is a uniform distribution over all permutations of $(0, - \one_{K-1})$, with $\mu = \frac{1}{K+1} \mu_1 + \frac{K}{K+1} \mu_0$.

  Let $\sigma \in \mathcal{D}(\sign)$ be a decentralized
  signaling mechanism. To make our notation slightly more compact, let $x^k_i(j) = 
  \sigma_k(j|i)$ be the probability that location~$k \in \comp$ sends signal $j \in \sign_k$ when it is in state $i \in \{ 0, \pm 1 \}$.
  Observe that $\sum_{j \in \sign_k} x^k_i(j) = 1$ for each $i$ and $k$. 

  Let $f \in \mathcal{F}(\sign)$ be an optimal strategy under
  $\sigma$; assume without loss of generality that $f$ is pure.

  Define
  $\exsign \coloneq \{ s \in \sign : \prob^\sigma(\hat{s} = s,
  \hat{\theta} = 0) > 0 \text{ and } f(0|s) = 0 \}$. Notice that if $s \in \exsign$, then there is a positive probability that signal $s$ will be sent when the state is a permutation of $(0, - \one_{K-1})$, and a customer using strategy $f$ will join the system upon receiving~$s$. For
  $k \in \comp$, define
  $\exsign^k \coloneq \{ s \in \exsign : f(k|s) = 1\}$. Note that
  $\exsign = \cup_{k \in \comp} \exsign^k$, and
  $\exsign^k \cap \exsign^\ell = \emptyset$ for $k \neq \ell$.

    Using \eqref{eq:thput}, the throughput under $\sigma$ and $f$ can be written as
    \begin{align}
      T(\sigma, f)
      &= \frac{1}{K+1} \sum_{\omega \in \Omega} \sum_{s \in \sign} \mu_1(\omega) \sigma(s | \omega)\sum_{k \in \comp} f(k|s)\notag\\
      &\quad \quad \quad + \frac{K}{K+1} \sum_{\omega \in \Omega} \sum_{s \in \sign} \mu_0(\omega) \sigma(s | \omega) \sum_{k \in \comp} f(k|s)\notag\\
       &\leq \frac{1}{K+1}  + \frac{K}{K+1} \sum_{\omega \in \Omega} \sum_{s \in \exsign}  \mu_0(\omega) \sigma(s| \omega) \notag\\
      &= \frac{1}{K+1} + \frac{K}{K+1}  \sum_{k \in \comp}  \sum_{s \in \exsign^k}  \sum_{\omega \in \Omega} \mu_0(\omega) \sigma(s| \omega) \notag\\
      &= \frac{1}{K+1} + \frac{K}{K+1}   \sum_{k \in \comp}  \sum_{s \in \exsign^k} \frac{1}{K} \cdot \sigma(s| \omega_k = 0 , \omega_{-k} = -\one_{K-1}) \notag\\
      &= \frac{1}{K+1} + \frac{1}{K+1}   \sum_{k \in \comp}  \sum_{s \in \exsign^k}  x^k_0(s_k) \prod_{\ell \neq k} x^\ell_{-1}(s_\ell)  \label{eq:reduced-throughput}
    \end{align}
    where, we have used 
    $\mu = \frac{1}{K+1} \mu_1 + \frac{K}{K+1} \mu_0$ in the first
    equality, and the definition of $\exsign$ in the inequality. In the second equality, we have interchanged the order of the summations and then used the fact that
    $\{ \exsign^k : k \in \comp\}$ is a partition of $\exsign$. To obtain
    the third equality, we use Part~1 of Lemma~\ref{lem:s1k-implications} in Section~\ref{sec:proofs} below, i.e., $\sigma(s|\omega) = 0$ if $s \in \exsign^k$ and 
    $\omega$ is such that $\omega_{k} = -1$. In addition, $\mu_0(w)=0$ for any $\omega$ that is not a permutation of $(0, -\one_{K-1})$. Thus, for each $k\in \comp$ and $s \in \exsign^k$, the summation over $\Omega$
    reduces to a single term corresponding to
    $\omega = (\omega_k, \omega_{-k}) = (0, -\one_{K-1})$. For this
    $\omega$, we have $\mu_0(\omega) = \frac{1}{K}$, giving us the third equality.

To bound the double sum in \eqref{eq:reduced-throughput}, define the sets  
 \begin{align*}
 \sign^k_+ &\coloneq \{ j \in \sign_k: x^k_1(j) > 0 ,  x^k_0(j) > 0 , x^k_{-1}(j) = 0\} \\
 \sign^k_{-} &\coloneq \{ j \in \sign_k : x^k_0(j) > 0, x^k_{-1}(j) > 0\}.
 \end{align*}

 Note that $\sign^k_{+}$ and $\sign^k_{-}$ depend on $x$; we hide this
 dependence for brevity. From
 Lemma~\ref{lem:s1k-implications} Part 3, we know that, if
 $s \in \exsign^k$, then $s_k \in \sign^k_+$ and
 $s_\ell \in \sign^\ell_{-}$ for all $\ell \neq k$. Letting
 $\sign_{{k-}} = \bigtimes\limits_{\ell \neq k} \sign^\ell_{-}$, we obtain
\begin{align*}
    \sum_{s \in \exsign^k}   x^k_0(s_k)  \prod_{\ell \neq k}  x^\ell_{-1}(s_\ell)
    &\leq \sum_{s_k \in \sign^k_+} x^k_0(s_k) \sum_{s_{-k} \in \sign_{{k-}}} \prod_{\ell \neq k} x^\ell_{-1}(s_\ell)\\
    &= \left( \sum_{a \in \sign^k_+} x^k_0(a) \right) \prod_{\ell \neq k} \left( \sum_{u \in \sign^\ell_{-}} x^\ell_{-1}(u) \right)\\
    &\leq  \sum_{a \in \sign^k_+} x^k_0(a),
\end{align*}
where in the final inequality, we use $\sum_{u \in \sign^\ell_{-}}
x^\ell_{-1}(u) \leq 1$ for all $\ell \in \comp$. Additionally, we have
\begin{align*}
    \sum_{s \in \exsign^k}   x^k_0(s_k)  \prod_{\ell \neq k}  x^\ell_{-1}(s_\ell)
    &\leq  \sum_{s \in \exsign^k}  x^k_1(s_k) \prod_{\ell \neq k} x^\ell_0(s_\ell)\\
    &\leq \sum_{s_k \in \sign^k_+} x^k_1(s_k) \sum_{s_{-k} \in \sign_{{k-}}} \prod_{\ell \neq k} x^\ell_0(s_\ell)\\
    &= \left(\sum_{a \in \sign^k_{+}} x^k_1(a)\right) \prod_{\ell \neq k} \left( \sum_{u \in \sign^\ell_{-}} x^\ell_0(u) \right)\\
    &\leq \prod_{\ell \neq k}\left( 1 - \sum_{u \in \sign^\ell_{+}} x^\ell_0(u) \right),  
\end{align*}
where we have used Lemma~\ref{lem:s1k-implications} Part 2 in the
first inequality. In the final inequality, we use the fact  that
$\sum_{u \in \sign^\ell_{+} \cup \sign^\ell_{-}} x^\ell_0(u) \leq 1$
for all $\ell \in \comp$.

Thus, taken together, we obtain for all $k \in \comp$
\begin{align*}
  \sum_{s \in \exsign^k}   x^k_0(s_k)  \prod_{\ell \neq k}  x^\ell_{-1}(s_\ell)
  &\leq \min\left\{ \sum_{a \in \sign^k_+} x^k_0(a), \prod_{\ell \neq k} \left( 1 - \sum_{a \in \sign^\ell_{+}} x^\ell_0(a) \right) \right\}\\
  &\leq \min\left\{ \sum_{a \in \sign^k_+} x^k_0(a), \left( 1 - \frac{1}{K-1} \sum_{\ell \neq k} \sum_{a \in \sign^\ell_{+}} x^\ell_0(a) \right)^{K-1} \right\}
\end{align*}
where we have used the AM-GM inequality in the last line.

Thus, we obtain
\begin{align}\label{eq:upperbound-asymmetric}
  \sum_{k \in \comp}  \sum_{s \in \exsign^k}   x^k_0(s_k)  \prod_{\ell \neq k}  x^\ell_{-1}(s_\ell) 
  \ & \ \leq \sum_{k \in \comp} \min\left\{ \sum_{a \in \sign^k_+} x^k_0(a), \left( 1 - \frac{1}{K-1} \sum_{\ell \neq k} \sum_{a \in \sign^\ell_{+}} x^\ell_0(a) \right)^{K-1} \right\}\notag \\
  \ & \ \leq \max_{u \in [0,1]^K} F(u),
\end{align}
where we define for all $u = (u_1, \dots, u_K) \in [0,1]^K$,
\begin{align}\label{eq:F}
  F(u) &\coloneq  \sum_{k \in \comp} \min\left\{ u_k, \left( 1 - \frac{1}{K-1} \sum_{\ell \neq k} u_\ell \right)^{K-1} \right\}\notag\\
  &= \sum_{k\in A(u)} u_k + \sum_{k \in A^c(u)} \left( 1 - \frac{1}{K-1}
    \sum_{\ell \neq k} u_\ell \right)^{K-1}, 
\end{align}
with
\begin{align} \label{eq:A}
A(u) &\coloneq \left\{ k \in \comp : u_k \leq \left( 1 - \frac{1}{K-1}
    \sum_{\ell \neq k} u_\ell \right)^{K-1} \right\}.
\end{align}

To complete the proof, we provide an upper bound on the maximization problem in  \eqref{eq:upperbound-asymmetric}. Lemma~\ref{lem:u-suffices} of Section~\ref{sec:proofs} shows
that to maximize $F$ over $u \in [0,1]^K$, it suffices to focus only on those $u \in [0,1]^K$ satisfying (a) $A(u) = \comp$, and (b) $u_k= u_\ell$
for all $k, \ell \in \comp$. Thus, assume $u_k = z$ for all $k$. Since $k \in A(u) = \comp$, we have
$z \leq (1 -z )^{K-1}$. For such $u$, we have $ F(u) = K z$. Hence
$F(u) \leq \max\{ Kz : z \leq (1-z)^{K-1}\} = Kz^*_K$, where $z^*_K \in [0,1]$ is the unique solution to
$z - (1-z)^{K-1}=0$.

Thus, we obtain from
\eqref{eq:upperbound-asymmetric} that
\begin{align*}
  \sum_{k \in \comp}  \sum_{s \in \exsign^k}   x^k_0(s_k)  \prod_{\ell \neq k}  x^\ell_{-1}(s_\ell)   \leq \max_{u \in [0,1]^K} F(u) \leq Kz^*_K.
\end{align*}

Substituting this bound into \eqref{eq:reduced-throughput}, we obtain
\begin{align*}
    T(\sigma, f) 
                \  \leq \ \frac{1}{K+1} + \frac{1}{K+1}   Kz^*_K \ 
  = \  \frac{1 + Kz^*_K}{1+K}.
\end{align*}  
It is straightforward to show that $z^*_K = \mathcal{O}(\log K/K)$, and
hence $T(\sigma, f) = \mathcal{O}(\log K/K)$.
\Halmos\endproof

An interesting direction for future research is to identify assumptions weaker than independence, but stronger than arbitrary dependence, under which stronger guarantees than those in Theorem~\ref{thm:correlated} can be obtained. In particular, in the above example, we note that the payoffs at different locations are positively correlated when $K \geq 4$. Whether better bounds can be achieved under negatively correlated payoffs is an interesting open question.

\section{Proofs and supporting results}
\label{sec:proofs}

\subsection{Supporting material for Section~\ref{sec:opt-dec}}

\medskip

\proof{Proof of Lemma~\ref{lem:decentralized_signal_space}.} Fix a decentralized signaling mechanism $\sigma \in \mathcal{D}(\fsign)$ where $\sigma(s|\omega) = \prod_{k \in \mathcal{K}} \sigma_k(s_k|\omega_k)$ for $s = (s_1,\dots,s_K) \in \fsign$ and $\omega=(\omega_1,\dots,\omega_K) \in \Omega$.  Let $\hat{s} = (\hat{s}_1, \dots, \hat{s}_K)$ be the random signal generated by $\sigma$; i.e., conditional on $\hat{\omega} = \omega$, we have $\hat{s} = s$ with probability $\sigma(s|\omega)$.

Define $\fsign_k^1 \coloneq \{ s_k \in \fsign_k : \expec^\sigma[h_k(\hat{\omega}_k) \ind\{ \hat{s}_k = s_k\}] \geq 0\}$, and $\fsign_k^0 \coloneq \fsign_k \setminus \fsign_k^1 = \{ s_k \in \fsign_k : \expec^\sigma[h_k(\hat{\omega}_k) \ind\{ \hat{s}_k = s_k\}] < 0\}$. Note that, for any $s_k \in \fsign_k^0$, we have $\expec^\sigma[h_k(\hat{\omega}_k) | \hat{s}_k = s_k] < 0$. For $u = (u_1, \dots, u_K) \in \uset$, let $\fsign^u \coloneq \fsign_1^{u_1} \times \fsign_2^{u_2} \times \dots \times \fsign_K^{u_K}$. Observe that $\{ \fsign^u : u \in \uset\}$ forms a partition of $\fsign$, i.e., $\cup_{u \in \uset} \fsign^u = \fsign$ and $\fsign^u \cap \fsign^v = \emptyset$ for $u \neq v \in \uset$.

Now, let $\psi \in \mathcal{D}(\uset)$ be the decentralized signaling mechanism defined by $\psi(u|\omega) \coloneq \prod_{k \in \mathcal{K}} \psi_k(u_k|\omega_k)$ 
for $u=(u_1,\dots,u_K) \in \uset = \{ 0,1\}^K$ where 
for each $k \in \comp$, we have 
\begin{align*}
    \psi_k( u_k|\omega_k)  \coloneq \sum_{s_k \in \fsign_k} \sigma_k(s_k | \omega_{k}) \ind\{ s_k \in \fsign_k^{u_k}\} \quad \text{for $u_k \in \{0, 1\}$.}
\end{align*}
In particular, this implies that for any $u \in \uset$, we have 
\begin{align}\label{eq:decent-u}
    \psi(u|\omega) = \sum_{s \in \fsign^u} \sigma(s|\omega) = \sum_{s \in \fsign^u} \prod_{k \in \comp} \sigma_k(s_k|\omega_k).
\end{align}
In words, we may think of $\psi$ generating a $\uset$-valued signal $\hat{u} = (\hat{u}_1, \dots, \hat{u}_K)$ as follows: each location $k\in \comp$, after learning the value of $\hat{\omega}_k=\omega_k$, independently generates an $\fsign_k$-valued signal $\hat{s}_k$ according to the distribution $\sigma_k(\cdot|\omega_k)$. Then, if $\hat{s}_k$ lies in the set $\fsign_k^1$, the location $k$ sends the signal $\hat{u}_k = 1$ to the customer, otherwise it sends the signal $\hat{u}_k = 0$. 

Note that $\hat{u}_k$ depends upon $\omega_k$, but not upon the other entries of $\omega$. Hence, the independence of $\hat{\omega}_1, \dots, \hat{\omega}_K$ implies that $\expec^{\psi}[h_k(\hat{\omega}_k) | \hat{u} = u] = \expec^{\psi}[h_k(\hat{\omega}_k) | \hat{u}_k = u_k]$ for any $u \in \uset$ with $\prob^\psi(\hat{u} = u) > 0$.

For any $u \in \uset\setminus\{\zero\}$ with $\prob^\psi(\hat{u} = u) > 0$, let $k$ be such that $u_k = 1$. Then, we have
\begin{align}
    \expec^\psi[ h_k(\hat{\omega}_k) |  \hat{u} = u] &= \expec^\psi[ h_k(\hat{\omega}_k) |  \hat{u}_k = 1] \notag    \\
    & = \frac{1}{\prob^\psi(\hat{u}_k = 1)}  \expec^\psi[ h_k(\hat{\omega}_k) \ind\{ \hat{u}_k = 1\}]\notag\\
    &= \frac{1}{\prob^\psi(\hat{u}_k = 1)} \sum_{s_k \in \fsign_k^1}   \expec^\sigma[ h_k(\hat{\omega}_k) \ind\{ \hat{s}_k = s_k\}]\notag\\
    &\geq 0,\label{eq:u-one}
\end{align}
where the third equality follows from the definition of $\psi$, and the inequality follows from the definition of $\fsign_k^1$. Thus, we conclude that upon receiving the signal $u$ under $\psi$, the customer (weakly) prefers joining the system to leaving without obtaining service: $\max_{\ell \in \comp} \expec^{\psi}[ h_{\ell}(\hat{\omega}_{\ell}) | \hat{u} = u] \geq 0$.  

Next, for any $u \in \uset\setminus\{\one\}$ with $\prob^\psi(\hat{u} = u) > 0$, let $k$ be such that $u_k = 0$. Then, similar to above,
\begin{align}
    \expec^\psi[h_k(\hat{\omega}_k) | \hat{u} = u] &=  \expec^{\psi}[ h_k(\hat{\omega}_k) | \hat{u}_k = 0]\notag\\
    &= \frac{1}{\prob^\psi(\hat{u}_k = 0)} \sum_{s_k \in \fsign_k^0}  \expec^\sigma[ h_k(\hat{\omega}_k) \ind\{ \hat{s}_k = s_k\}] \notag\\
    &< 0.\label{eq:u-zero}
\end{align}
Thus, we conclude that upon receiving the signal $u$ under the mechanism $\psi$, the customer (strictly) prefers leaving the system altogether over joining location $k$. This further implies that $\max \{ \max_{\ell \in \comp,\ell \neq k} \expec^{\psi}[ h_{\ell}(\hat{\omega}_{\ell}) | \hat{u} = u]  , 0\} > \expec^{\psi}[ h_k(\hat{\omega}_k) | \hat{u} = u]$.

Taken together, \eqref{eq:u-one} and \eqref{eq:u-zero} imply that under the signaling mechanism $\psi$, there exists an optimal strategy $\hat{f}$ such that (i) for any $u \in \uset\setminus\{\zero\}$, the customer always joins one of the locations, i.e., $\hat{f}(0|u) = 0$, and (ii) for any $u \in \uset\setminus\{\one\}$, the customer never joins a location $k \in \comp$ with $u_k = 0$, i.e., $\hat{f}(k|u) = 0$. In other words, we conclude that there exists a strategy $\hat{f} \in \mathcal{F}_d$ that is optimal under $\psi$.

Finally, let $f \in \mathcal{F}(\fsign)$ be an optimal strategy under the mechanism $\sigma$. For $s \in \fsign^\zero$, we have $\expec^\sigma[h_k(\hat{\omega}_k) | \hat{s} = s] = \expec^\sigma[h_k(\hat{\omega}_k) | \hat{s}_k = s_k] < 0$ for all $k \in \comp$. Since $f$ is optimal, this implies $f(k|s) = 0$ for all $k \in \comp$ and $s \in \fsign^\zero$. Consequently, $f(0|s) =1$. Thus, we obtain, for any $\hat{f} \in \mathcal{F}_d$, 
\begin{align*}
    T(\psi, \hat{f}) &= \sum_{\omega \in \Omega } \sum_{u \in \uset}  \mu(\omega)  \psi(u|\omega) \sum_{k \in \comp} \hat{f}(k|u)\\
    &= \sum_{\omega \in \Omega } \sum_{u \in \uset\setminus\{\zero\}}  \mu(\omega)  \psi(u|\omega) \\
    &= \sum_{\omega \in \Omega } \sum_{u \in \uset\setminus\{\zero\}}  \mu(\omega)  \sum_{s \in \fsign^u} \sigma(s|\omega) \\
    &\geq \sum_{\omega \in \Omega } \sum_{u \in \uset}  \mu(\omega)  \sum_{s \in \fsign^u} \sigma(s|\omega) \sum_{k \in \comp} f(k|s) \\
    &= T(\sigma, f).
\end{align*}
Here, the second equality follows from the fact that $\sum_{k \in \comp} \hat{f}(k|u) = \ind\{u \neq \zero\}$ for all $u \in \uset$. The third equality follows from \eqref{eq:decent-u}. The inequality follows from the fact that $\sum_{k \in \comp} f(k|s) \leq 1$ for $s \in \fsign$, and $\sum_{k \in \comp} f(k|s) = 0$ for $s \in \fsign^\zero$. The last equality follows because $\cup_{u \in \uset} \fsign^u = \fsign$. 
 \Halmos\endproof

\proof{Proof of Lemma~\ref{lem:decentralized-obedience}.} Consider any $\sigma \in \mathcal{D}(\uset)$ and let $\hat{s} = (\hat{s}_1, \dots, \hat{s}_K)$ be a random signal with conditional distribution $\sigma(s|\omega)$. We begin by proving the forward implication. 

Suppose there exists an $f \in \mathcal{F}_d$ that satisfies \eqref{eq:optimal-strategy} under $\sigma$. We split our analysis into two cases.

\textsf{Case 1: $\prob^{\sigma}(\hat{s} = \zero)  = 0$}. In this case, we will show that condition~(I) holds. From the independence of the location states and the signals, we conclude that there exists at least one location $k \in \comp$ with $\prob^{\sigma}(\hat{s}_k = 1) = 1$. Let $\compn_1 \coloneq \{ k \in \comp : \prob^{\sigma}(\hat{s}_k = 1) = 1 \} \neq \emptyset$. Observe that for all $k \in \compn_1$, we have for all $\omega_k \in \Omega_k$,  if $\mu_k(\omega_k) > 0$ then $\sigma_k(0| \omega_k)  =0$, and hence $\mu_k(\omega_k)\sigma_k(0|\omega_k) = 0$.

Let $z \coloneq \sum_{k \in \compn_1} e_k \in \{0, 1\}^K$. By definition of $\compn_1$, we have $\prob^{\sigma}( \hat{s} = z) > 0$. Because $f \in \mathcal{F}_d$, there exists a $k \in \compn_1$ such that $f(k|z) > 0$. Furthermore, since $f$ is optimal under $\sigma$, we obtain that 
\begin{align}
    \expec^\sigma\left[ h_k(\hat{\omega}_k) | \hat{s} = z \right] \geq \max\left\{ 0, \max_{\ell \neq k} \expec^\sigma\left[ h_\ell(\hat{\omega}_\ell) | \hat{s} = z \right] \right\}.
    \label{eq:cond-exp-h}
\end{align}
Using the independence of $(\hat{\omega}_k, \hat{s}_k)$ and $\hat{s}_{-k}$, together with the fact that $\prob^{\sigma}( \hat{s}_{-k} = z_{-k}) \geq \prob^\sigma( \hat{s} = z) > 0$, we obtain $\sum_{\omega_k \in \Omega_k} \mu_k(\omega_k) h_k(\omega_k) \geq 0$. 

For any $\ell \in \compn_1$, both sides of the final inequality in condition (I) are equal to zero, and hence the inequality holds trivially. For $\ell \notin \compn_1$, we have $z_\ell = 0$. Using the independence of $(\hat{\omega}_k, \hat{s}_k)$, $(\hat{\omega}_{\ell}, \hat{s}_\ell)$ and $(\hat{\omega}_{-(k, \ell)}, \hat{s}_{-(k, \ell)})$, together with the fact that $\prob^\sigma( \hat{s}_{-(k, \ell)} = z_{-(k, \ell)}) \geq \prob^\sigma( \hat{s} = z) > 0$, we obtain from \eqref{eq:cond-exp-h} that
\begin{align*}
    \expec^\sigma\left[ h_k(\hat{\omega}_k) \ind\{\hat{s}_k = 1, \hat{s}_\ell = 0\}\right] \geq  \expec^\sigma\left[ h_\ell(\hat{\omega}_\ell) \ind\{\hat{s}_k = 1, \hat{s}_\ell = 0\}\right].
\end{align*}
Expanding the expectations, and noticing that $\sigma_k(1|\omega_k) = 1$ for all $\mu_k(\omega_k) > 0$, we obtain the final inequality in condition (I).

\textsf{Case 2: $\prob^{\sigma}(\hat{s} = \zero) > 0$}. In this case, we will show that condition~(II) holds. Because $f \in \mathcal{F}_d$, we have $f(0|\zero) = 1$. By the optimality of $f$, we obtain $\expec^\sigma\left[ h_k(\hat{\omega}_k) | \hat{s} = \zero\right] \leq 0$ for all $k \in \comp$. Using the independence of $(\hat{\omega}_k, \hat{s}_k)$ and $\hat{s}_{-k}$, together with the fact that $\prob^{\sigma}( \hat{s}_{-k} = \zero) \geq \prob^\sigma( \hat{s} = \zero) > 0$, we conclude that $\sum_{\omega_k \in \Omega_k} \mu_k(\omega_k) \sigma_k(0|\omega_k) h_k(\omega_k) \leq 0$ for all $k \in \comp$.

Let $\compn_0 \coloneq \{ k \in \comp : \prob^\sigma( \hat{s}_k = 0) = 1\}$. Then, for any $k \in \compn_0$, we have $\sigma_k(0|\omega_k) = 1- \sigma_k(1|\omega_k) = 1$ for all $\omega_k \in \Omega_k$ with $\mu_k(\omega_k) > 0$. Thus, we obtain $\sum_{\omega_k \in \Omega_k} \mu_k(\omega_k) \sigma_k(1|\omega_k) h_k(\omega_k) = 0$ for all $k \in \compn_0$. For any $k \notin \compn_0$, we have $\prob^\sigma(\hat{s}_k = 0) \in (0,1)$, and hence $\prob^\sigma( \hat{s} = e_k) > 0$. For any such $k \notin \compn_0$, from the optimality of $f$ and the fact that $f(k| e_k) = 1$, we conclude that $\expec^\sigma\left[ h_k(\hat{\omega}_k) | \hat{s} = e_k\right] \geq 0$. Once again, using the independence of $(\hat{\omega}_k, \hat{s}_k)$ and $\hat{s}_{-k}$, together with the fact that $\prob^{\sigma}( \hat{s}_{-k} = \zero) \geq \prob^\sigma( \hat{s} = e_k) > 0$, we conclude that $\sum_{\omega_k \in \Omega_k} \mu_k(\omega_k) \sigma_k(1|\omega_k) h_k(\omega_k) \geq 0$ for all $k \notin \compn_0$. This establishes that condition~(II) holds.

To prove the converse, we will first construct, for any $\sigma \in \mathcal{D}(\uset)$, a strategy $f \in \mathcal{F}_d$ and then show that it is optimal under $\sigma$ for either of the two conditions in the lemma statement. Toward that end, fix $\sigma \in \mathcal{D}(\uset)$. For each $k \in \comp$ and $u_k \in \{0,1\}$, define 
\begin{align*}
    H_k^\sigma(u_k) &\coloneq \begin{cases}
        \expec^\sigma\left[ h_k(\hat{\omega}_k) | \hat{s}_k = u_k\right] & \text{if $\prob^\sigma( \hat{s}_k = u_k) > 0$;}\\
        -\infty & \text{otherwise.}
    \end{cases}
\end{align*}
For $u \in \uset$, let $H^\sigma(u) \coloneq \max_{k \in \comp} H_k^\sigma(u_k)$
and $A^\sigma(u) \coloneq \arg \max_{k \in \comp} H_k^\sigma(u_k)$. Define 
\begin{align*}
    k^*(u) = \begin{cases}  0 & \text{if $u = \zero$;}\\
    \min\{ k \in A^\sigma(u) : u_k = 1 \} & \text{if $\{ k \in A^\sigma(u) : u_k = 1\} \neq \emptyset$;}\\
    \min \{ k \in \comp : u_k = 1 \} & \text{if $\{ k \in A^\sigma(u) : u_k = 1\} = \emptyset$ and $u \neq \zero$.} 
    \end{cases}
\end{align*}

Finally, define $f \in \mathcal{F}(\uset)$ as follows: for any $u \in \uset$, we let $f(k^*(u) | u) = 1$. It is straightforward to see that $f \in \mathcal{F}_d$. We will now show that $f$ indeed is optimal under $\sigma$ when either of the two conditions in the lemma statement are met.

Suppose first that condition (I) holds, and let $\ell \in \comp$ be the location for which the statements in the condition hold. Note that these conditions imply that $H_\ell^\sigma(0) = -\infty$, $H_\ell^\sigma(1) \geq 0$ and $H_\ell^\sigma(1) \geq H_k^\sigma(0)$ for all $k \neq \ell \in \comp$. 

For any $u \in \uset$ with $u_\ell = 1$, since $H_\ell^\sigma(1) \geq H_k^\sigma(0)$ for all $k \neq \ell \in \comp$, we obtain that $\{ j \in A^\sigma(u) : u_j = 1\} \neq \emptyset$.
Thus, we get that $j = k^*(u)$ satisfies $H_j^\sigma(u_j) =  H^\sigma(u) \geq H^\sigma_\ell(u_\ell) \geq 0$, and hence $f$ selects an optimal location to join. Since for any $u \in \uset$ with $\prob^\sigma(\hat{s} = u) > 0$, we have $u_\ell = 1$, we conclude that $f$ is indeed optimal under $\sigma$.

Next, suppose that condition~(II) holds, i.e.,  $\sigma \in \mathcal{D}(\uset)$ satisfies~\eqref{eq:sys_mod_con1} and~\eqref{eq:sys_mod_con2}. These conditions imply that for any $u \neq \zero$ with $\prob^\sigma(\hat{s} = u) > 0$, we have $\max_{k \in \comp : u_k = 1} H_k^\sigma(1) \geq 0 \geq \max_{k \in \comp : u_k = 0} H_k^\sigma(0)$, and hence $f$ indeed selects an optimal location $k^*(u)$ to join for any such $u$. On the other hand, if $\prob^\sigma(\hat{s} = \zero) > 0$, then we have $\max_{k \in \comp} H_k^\sigma(0) \leq 0$, and hence, we again obtain that $f$ selects an optimal action because $k^*(\zero) = 0$. So, $f$ is optimal under $\sigma$. \Halmos\endproof

\subsection{Supporting material for Sections~\ref{sec:cost-dec} and~\ref{sec:correlated}}

\medskip

\begin{lemma}\label{lem:series_algebra}
For any $x_1,x_2,\dots,x_K \in [0,1]$ with $\sum_{k=1}^K x_k \in [0,1]$, we have
\begin{align*}
 1-\prod_{k = 1}^K(1-x_{k}) \geq \left[ 1-\left(1-\frac{1}{K}\right)^K \right] \cdot 
 \sum_{k = 1}^K x_{k}.
\end{align*}
\end{lemma}

\proof{Proof.} For $y_1,\dots,y_K \geq 0$, the AM-GM inequality implies ${(\frac{1}{K} \sum_{k =1}^K y_k)}^K \geq \prod_{k =1}^K y_k$. Taking $y_k = 1 - x_k$  yields $(1-\bar{x})^K \geq \prod_{k=1}^K ( 1- x_k)$ where $\bar{x} = \frac{1}{K} \sum_{k =1}^K x_k$.  It follows that 
\begin{align*}
    1 - \prod_{k=1}^K (1 - x_k) \geq 1 - (1 - \bar{x})^K.
\end{align*}
Note that $\bar{x} \in [0,\frac{1}{K}]$. If $\bar{x} = 0$, the inequality in the lemma statement holds trivially. Hence, assume $\bar{x} > 0$.  Letting $F(x) \coloneq \frac{1-(1-x)^K}{K x}$, we obtain
\begin{align*}
\frac{1-\prod_{k=1}^K(1-x_{k})}{\sum_{k=1}^K x_{k}} \geq \frac{1-(1-\bar{x})^K}{K \bar{x}} = F(\bar{x}). 
\end{align*}
The function $F$ is decreasing on $(0,1/K]$. Thus, $F(\bar{x}) \geq F(1/K) = 1-{\left(1-1/K\right)}^K$, and hence
\begin{align*}
    1 - \prod_{k=1}^K (1 - x_k) \geq 1 - (1 - \bar{x})^K \geq K \bar{x} F(\bar{x}) \geq \left[ 1-{\left(1-\frac{1}{K}\right)}^K \right] \cdot \sum_{k=1}^K x_k.
\end{align*}
This proves the lemma.
\Halmos\endproof

\begin{lemma}\label{lem:s1k-implications} Suppose that $\sigma \in \mathcal{D}(\sign)$ is a decentralized signaling mechanism for the system analyzed in Theorem~\ref{thm:correlated-upper-bound}. Let $x_{i}^k(j) = \sigma_k(j|i)$ for $i \in \{0, \pm 1\}$, $j \in \sign_k$, and $k \in \comp$. Let $\exsign^k$ be as defined in the proof of Theorem~\ref{thm:correlated-upper-bound}. 
For all $k \in \comp$ and all $s \in \exsign^k$, we
have the following:
\begin{enumerate}
    \item For any $\omega \in \Omega$ with $\omega_{k} = -1$, we have
    $\sigma(s| \omega) = 0$. In particular, for all $\ell \neq k$,
    \begin{align}\label{eq:minus-one-sigma-2}
      x^k_{-1}(s_k) x^\ell_0(s_\ell) \prod_{m \neq k, \ell} x^m_{-1}(s_m) = 0.
    \end{align}

  \item We have
    \begin{align}
        \label{eq:one-zero-sigma}
      x^k_0(s_k) \prod_{\ell \neq k} x^\ell_{-1}(s_\ell)  \ \leq \  x^k_1(s_k) \prod_{\ell \neq k} x^\ell_0(s_\ell).
    \end{align}

  \item Finally, we have
      \textup{(a)} $x^k_1(s_k)> 0$, \textup{(b)} $x^k_0(s_k) > 0$, \textup{(c)} $x^k_{-1}(s_k) = 0$, and also \textup{(d)} $x^\ell_0(s_\ell) > 0$ for all $\ell \neq k$, \textup{(e)} $x^\ell_{-1}(s_\ell) > 0$
    for all $\ell \neq k$.
\end{enumerate}
\end{lemma}

  \proof{Proof.} Fix $k \in \comp$ and $s = (s_1, \dots, s_K) \in \exsign^k$.

    \textsf{Part 1.} Since $f$ is optimal and $f(k|s) = 1$, we obtain
    that $\expec^\sigma[ h_k(\hat{\omega}_k) | \hat{s} = s] \geq 0$.
    For this to hold, it must be that
    $\prob^\sigma( \hat{\omega}_k = -1 | \hat{s} = s) = 0$, which
    implies that
    $\prob^\sigma( \hat{s} = s | \hat{\omega}_k = -1 ) = 0$. This
    implies that $\sigma(s|\omega) = 0$ for all $\omega$ with
    $\omega_k = -1$.
    For any $\ell \neq k$, and $\omega$ with
    $\omega_{\ell} = 0$ and $\omega_{-\ell} = -\one_{K-1}$,
    $\sigma(s|\omega) =0$ implies~\eqref{eq:minus-one-sigma-2}.

    \textsf{Part 2.} Because $f(k|s) = 1$, we have
    $\expec^\sigma[ h_k(\hat{\omega}_k) | \hat{s} = s] \geq 0$. This
    implies
    \begin{align*}
      \prob^\sigma( \hat{\omega}_k = 1 | \hat{s} = s) \cdot K + \prob^\sigma( \hat{\omega}_k = 0 | \hat{s} = s) (-1) \geq 0.
    \end{align*}
    Note that
    $\prob^\sigma( \hat{\omega}_k =1 | \hat{s} = s) = \prob^\sigma (
    \hat{\omega}_{-k} = \zero_{K-1} , \hat{\omega}_k = 1 | \hat{s} =
    s)$, and
    $\prob^\sigma( \hat{\omega}_k = 0 | \hat{s} = s) = \prob^\sigma(
    \hat{\omega}_{-k} = -\one_{K-1}, \hat{\omega}_k = 0 | \hat{s} = s)
    + \sum_{\ell \neq k} \prob^\sigma( \hat{\omega}_{-\ell} =
    \zero_{K-1}, \hat{\omega}_\ell = 1 | \hat{s} = s) $. Thus, we
    have
    \begin{align*}
      \prob^\sigma( \hat{\omega}_{-k} = \zero_{K-1}, \hat{\omega}_k = 1 | \hat{s} = s) \cdot K
      &+       \prob^\sigma( \hat{\omega}_{-k} = -\one_{K-1}, \hat{\omega}_k = 0 | \hat{s} = s) (-1)\\
      &+ \sum_{\ell \neq k} \prob^\sigma( \hat{\omega}_{-\ell} =
    \zero_{K-1}, \hat{\omega}_\ell = 1 | \hat{s}
    = s) (-1) \geq 0.
    \end{align*}    
which implies
    \begin{align*}
      \prob^\sigma( \hat{\omega}_{-k} = \zero_{K-1}, \hat{\omega}_k = 1 | \hat{s} = s) \cdot K 
       +       \prob^\sigma( \hat{\omega}_{-k} = -\one_{K-1}, \hat{\omega}_k = 0 | \hat{s} = s) (-1) \ \geq \ 0.
    \end{align*}  
    Using the fact that $\prob^\sigma(\hat{s} = s) > 0$ and Bayes' rule, we obtain
    \begin{align*}
      &\prob^\sigma( \hat{s} = s | \hat{\omega}_{-k} = \zero_{K-1}, \hat{\omega}_k = 1 ) \cdot \prob(\hat{\omega}_{-k} = \zero_{K-1}, \hat{\omega}_k = 1)  \cdot K\\
      &+       \prob^\sigma( \hat{s} = s | \hat{\omega}_{-k} = -\one_{K-1}, \hat{\omega}_k = 0 ) \cdot \prob(\hat{\omega}_{-k} = -\one_{K-1}, \hat{\omega}_k = 0) \cdot (-1) \geq 0.
    \end{align*}
    From the definition of $\mu$, we have
    \begin{align*}
      &\prob^\sigma( \hat{s} = s | \hat{\omega}_{-k} = \zero_{K-1}, \hat{\omega}_k = 1 ) \cdot \frac{1}{K(K+1)}  \cdot K\\
      &+       \prob^\sigma( \hat{s} = s | \hat{\omega}_{-k} = -\one_{K-1}, \hat{\omega}_k = 0 ) \cdot \frac{K}{K(K+1)}\cdot (-1) \ge 0, 
    \end{align*}
    which is equivalent to
      $\prob^\sigma( \hat{s} = s | \hat{\omega}_{-k} = -\one_{K-1}, \hat{\omega}_k  = 0 ) 
      \leq  \prob^\sigma( \hat{s} = s | \hat{\omega}_{-k} = \zero_{K-1}, \hat{\omega}_k = 1 )$.
    The statement in Part 2 follows by expanding each term in this inequality.

    \textsf{Part 3.} 
    Since
    $\prob^\sigma( \hat{s} = s , \hat{\theta} = 0) > 0$, we have
    \begin{align*}
      0 &<   \sum_{\omega \in \Omega} \mu_0(\omega) \prob^\sigma(\hat{s} = s | \hat{\omega} = \omega) \\
        &= \sum_{\ell \in \comp} \frac{1}{K} \prob^\sigma(\hat{s} = s | (\hat{\omega}_\ell, \omega_{-\ell})  = (0, -\one_{K-1})) \\
        &= \frac{1}{K} \sum_{\ell \in \comp} x^\ell_0(s_\ell) \prod_{m \neq \ell} x^m_{-1}(s_m) \\
        &= \frac{1}{K} x^k_0(s_k) \prod_{\ell \neq k} x^\ell_{-1}(s_\ell) + \frac{1}{K} \sum_{\ell \neq k } x^\ell_0(s_\ell) x^k_{-1}(s_k) \prod_{m \neq \ell,k} x^m_{-1}(s_m) \\
      &= \frac{1}{K} x^k_0(s_k) \prod_{\ell \neq k} x^\ell_{-1}(s_\ell),
    \end{align*} 
    where the final equality follows from
    \eqref{eq:minus-one-sigma-2}. This yields 
    \begin{align}\label{eq:fsign-s}
      x^k_0(s_k) \prod_{\ell \neq k} x^\ell_{-1}(s_\ell) > 0.
    \end{align}
    
    Comparing \eqref{eq:minus-one-sigma-2} and
    \eqref{eq:fsign-s}, we obtain,
    \begin{align*}
      \prod_{m \neq \ell, k} x^m_{-1}(s_m) &> 0 \quad \text{for all
                                           $\ell \neq k$,}\\
      x^k_0(s_k)x^\ell_{-1}(s_\ell) &> 0 \quad \text{for all $\ell \neq k$,}\\
      x^k_{-1}(s_k)x^\ell_0(s_\ell) &= 0 \quad \text{for all $\ell \neq k$.}
    \end{align*}
    From this, we conclude that $x^k_0(s_k) > 0$ and that $x^\ell_{-1}(s_\ell) >0$ for all $\ell \neq k$. This establishes (b) and (e) in Part 3. From the above we also have for each $\ell \neq k$ that
    \begin{align}
      x^\ell_0(s_\ell) = 0 \ \ \text{ or } \ \ x^k_{-1}(s_k) = 0 .\label{eq:alternative-sigma}
    \end{align}
    From \eqref{eq:fsign-s}, we know the 
    left side of \eqref{eq:one-zero-sigma} is strictly positive, and hence so is the
    right side. This implies that $x^k_1(s_k) > 0$ and $x^\ell_0(s_\ell) > 0$ for
    all $\ell \neq k$. Therefore, (a) and (d) hold. In turn, from \eqref{eq:alternative-sigma}, we
    conclude that $x^k_{-1}(s_k) = 0$, which proves (c).
\Halmos\endproof

\begin{lemma}\label{lem:u-suffices} 
Consider the problem $ \max_{u \in [0,1]^K} F(u)$ where $F(\cdot)$ is defined in \eqref{eq:F}. It suffices to consider only $u \in
  [0,1]^K$ satisfying \textup{\textsf{(a)}} $A(u) = \comp$ where $A(\cdot)$ is defined in \eqref{eq:A}; and \textup{\textsf{(b)}} $u_k = u_\ell$ for all
  $k, \ell \in \comp$.
\end{lemma}
\proof{Proof.}
  \textsf{Part (a).} For $u \in [0,1]^K$, if $|A^c(u)| \geq 2$, then we have for all $k \in A^c(u)$,
\begin{align*}
  u_k > \left(1 - \frac{1}{K-1} \sum_{\ell \neq k} u_\ell\right)^{K-1} \geq 0,
\end{align*}
and
$F(u) = \sum_{k \in A(u)} u_k + \sum_{k \in A^c(u)} \left(1 - \frac{1}{K-1} \sum_{\ell \neq k}
  u_\ell\right)^{K-1}$. Letting $\one_{A^c(u)}$ be the vector with $1$ at each component $k \in A^c(u)$ and $0$ otherwise, it is straightforward to see that $v := u - \epsilon \one_{A^c(u)}$ is in $[0,1]^K$ for small enough $\epsilon>0$, and $A(v) = A(u)$. Furthermore, we have
 $F(v) = F(u - \epsilon \one_{A^c(u)}) > F(u)$ for small enough $\epsilon >0$, since each term in $F$ corresponding to a $k \in A^c(v)$ increases (while those corresponding to $k \in A(u)$ remain the same). Hence, such $u$ cannot be optimal.

Next, suppose $u \in [0,1]^K$ is such that $|A^c(u)| = 1$. Without loss of generality, assume $A(u) = \{1, \dots, K-1\}$ and $A^c(u) = \{K\}$. Then we have
\begin{align*}
u_k &\leq \left( 1 - \frac{1}{K-1} \sum_{\ell \neq k} u_\ell\right)^{K-1} \quad \text{for $k < K$,}\\
u_K  &> \left( 1 - \frac{1}{K-1} \sum_{\ell \neq K} u_\ell\right)^{K-1}.
\end{align*}
Observe that $u_K > 0$, and hence $u_k < 1$ for all $k < K$. Define $v \in [0,1]^K$ as follows:
\begin{align*}
    v_k \coloneq \begin{cases} u_k + \epsilon_1 & \text{if $k < K$;}\\
    u_K - \epsilon_2 & \text{if $k = K$.}
    \end{cases}
\end{align*}
for small positive $\{\epsilon_i : i =1,2\}$. Since, $u_K > 0$ and $u_k < 1$ for all $k < K$, we note that indeed $v \in [0,1]^K$ for all small enough $\{ \epsilon_i : i =1,2\}$. 

We have for $k < K$,
\begin{align*}
    v_k - \left( 1 - \frac{1}{K-1} \sum_{\ell \neq k} v_\ell\right)^{K-1} 
       = \ \ u_k + \epsilon_1 - \left( 1 - \frac{1}{K-1} \sum_{\ell \neq k} u_\ell - \frac{K-2}{K-1}\epsilon_1  + \frac{1}{K-1}\epsilon_2\right)^{K-1} 
     \leq \ \ 0,
\end{align*}
for small enough $\epsilon_i > 0$ with $\epsilon_2 > (K-2) \epsilon_1$. Thus, $k \in A(v)$. On the other hand, from continuity arguments, we have
\begin{align*}
    v_K - \left( 1 - \frac{1}{K-1} \sum_{\ell \neq K} v_\ell\right)^{K-1} 
    &  = \ \ u_K - \epsilon_2 - \left( 1 - \frac{1}{K-1} \sum_{\ell \neq K} u_\ell - \epsilon_1\right)^{K-1} 
     > \ \ 0,
\end{align*}
for small enough $\epsilon_i > 0$. Thus, we obtain $K \in A^c(v)$. Thus, we conclude that $A(u) = A(v)$. Finally, for small enough $\epsilon_1$, we have
\begin{align*}
    F(v) &= \sum_{k \in A(v)} v_k + \sum_{k \in A^c(v)} \left( 1 - \frac{1}{K-1} \sum_{\ell \neq k} v_\ell\right)^{K-1} \\
    &= \sum_{k =1}^{K-1} u_k + (K-1) \epsilon_1 + \left( 1 - \frac{1}{K-1} \sum_{\ell \neq K} u_\ell - \epsilon_1\right)^{K-1} \\
    &> \sum_{k =1}^{K-1} u_k +  \left( 1 - \frac{1}{K-1} \sum_{\ell \neq K} u_\ell \right)^{K-1}\\ 
    &= F(u).
\end{align*}
Here, in the inequality, we have used the fact that $\frac{1}{K-1}\sum_{\ell \neq K} u_\ell < 1$. The above implies that any $u \in [0,1]^K$ with $|A^c(u)| = 1$ cannot be optimal either. Thus, we conclude that it suffices to consider $u \in [0,1]^K$ with $|A^c(u)| = 0$, i.e.,  $A(u) = \comp$.

\textsf{Part (b).} Consider any $u \in [0,1]^K$ with $A(u) = \comp$. Let $u_{\min} \coloneq \min_{k \in \comp} u_k$ and $u_{\max} \coloneq \max_{k \in \comp} u_k$. If $u_{\min} = u_{\max}$, then $u$ already meets the condition \textsf{(b)}. So, let $u_{\min} < u_{\max}$. Then,  
\begin{align*}
    u_k &\leq \left( 1 - \frac{1}{K-1}   \sum_{\ell \neq k} u_\ell \right)^{K-1} \quad \text{for all $k\in \comp$.}
\end{align*}
Define the function $G$ as follows:
\begin{align*}
    G(x) &\coloneq x - \left( 1 - \frac{1}{K-1} \sum_{\ell \in \comp} u_\ell  + \frac{x}{K-1}\right)^{K-1}
\end{align*}
Note that $G(u_k)\leq 0$ for all $k \in \comp$. For all $x \in [u_{\min}, u_{\max}]$, we have
\begin{align*}
    G'(x) = 1 - \left( 1 - \frac{1}{K-1} \sum_{\ell \in \comp} u_\ell + \frac{x}{K-1}\right)^{K-2} \geq 0.
\end{align*}
Since $G$ is increasing over $x \in [u_{\min}, u_{\max}]$ and $G(u_{\max}) \leq 0$, we conclude that $G(x) \leq 0$ for all $x \in [u_{\min}, u_{\max}]$.

Define $v \coloneq \left(\frac{1}{K}  \sum_{\ell \in K} u_\ell\right) \one_K$. We have  $v_k \in [u_{\min}, u_{\max}]$ for all $k \in \comp$, and hence 
\begin{align*}
    0 \geq  G(v_k) = v_k - \left( 1 - \frac{1}{K-1} \sum_{\ell \in \comp} u_\ell  + \frac{v_k}{K-1}\right)^{K-1}. 
\end{align*} 
Since $\sum_{\ell \in \comp} u_\ell = \sum_{\ell \in \comp} v_\ell$, this implies $k \in A(v)$.  Thus, 
$A(v) = A(u) = \comp$. It is easy to see that $F(v) = F(u)$. Hence,
it suffices to consider $u \in [0,1]^K$ with $u_k = u_\ell$ for all
$k, \ell \in \comp$.
\Halmos\endproof

\bibliographystyle{abbrvnat}
\bibliography{references}

\end{document}